\documentclass[prb,twocolumn, showpacs,aps,superscriptaddress,floatfix]{revtex4-2}
\usepackage{amsmath}
\usepackage{amssymb}
\usepackage{amsfonts}
\usepackage{bm}
\usepackage[dvips]{graphicx}
\usepackage{color}
\usepackage{ulem}
\usepackage[unicode, pdftex]{hyperref}
\usepackage{verbatim}% Long comments
\begin{document}

\title{Anomalous Josephson Hall effect in doped topological insulators with the nematic superconductivity}

\author{R.S. Akzyanov}
\affiliation{Dukhov Research Institute of Automatics, Moscow, 127055 Russia}
\affiliation{Moscow Institute of Physics and Technology, Dolgoprudny,
Moscow Region, 141700 Russia}
\affiliation{Institute for Theoretical and Applied Electrodynamics, Russian
Academy of Sciences, Moscow, 125412 Russia}
\author{A.L. Rakhmanov}
\affiliation{Dukhov Research Institute of Automatics, Moscow, 127055 Russia}
\affiliation{Institute for Theoretical and Applied Electrodynamics, Russian
Academy of Sciences, Moscow, 125412 Russia}

\begin{abstract}
We study the physics of the Josephson effect in nematic superconductors with $E_u$ odd parity in the Ginzburg-Landau approach. Two-component vector superconducting order parameter makes this effect rather unusual. We get that the Meissner kernel has off-diagonal components. We derive current-phase relations for different configurations of the junction, crystallographic axes of the sample, and nematicity direction. We show that an anomalous Josephson Hall effect can be observed in such a system without any magnetization. That is, for definite orientations of the junction and crystal axes, a component of the Josephson current along the junction is induced by the order parameter phase difference across the contact. We also calculate the magnetic field dependence of the maximum current through the junction. We find that the period of the Fraunhofer oscillations of the maximum Josephson current depends on the geometry of the junction, direction of the magnetic field, and nematicity vector. 
\end{abstract}

\maketitle

\section{Introduction}

The Josephson effect is a general feature of superconducting phenomena~\cite{barone1982, Tinkham2004, Likharev1979}. It manifests itself as a generation of the supercurrent between two superconducting pieces that are separated by a thin barrier. The value of this current is controlled by the external gating that induces phase difference between superconductors. This effect is not only of importance for many superconducting devices but also provides important information about the physical nature of the superconducting order. The discovery of topological superconductivity opens a new prospect in the Josephson physics~\cite{tafuri2019}. In particular, the study of the Josephson currents in superconductor-ferromagnetic (SF) heterostructures gives rise to the observation of a set of new physical phenomena and the development of a new family of superconducting devices~\cite{Melnikov2021}. An important property of the SF systems is the possibility to govern the Josephson current by fine-tuning of magnetization of the F layer. 

Recently, a significant interest arises to a ``transverse'' Josephson currents that flows along the junction~\cite{Wang2011,Vedyayev2013,Yokoyama2015,MatosAbiague2015,Dang2015,Mironov2017,Malshukov2019, Costa2020, Maistrenko2021}. 
This phenomenon is called Josephson Hall effect, anomalous Josephson Hall effect, or tunnel Josephson Hall effect by an analogy with the anomalous Hall effect in the normal state~\cite{Nagaosa2010}. The Josephson Hall effect arises due to non-trivial interplay between the magnetization and spin-orbit interaction~\cite{Vedyayev2013,MatosAbiague2015,Dang2015,Mironov2017,Malshukov2019,Costa2020,Maistrenko2021} or unconventional superconductivity~\cite{Wang2011,Yokoyama2015}. Such a transverse Josephson current can be comparable with the usual ``longitudinal'' Josephson current~\cite{Maistrenko2021}.

Experiments with doped topological insulators such as Cu$_x$Bi$_2$Se$_3$~\cite{Hor2010,Sasaki2011,Kirzhner2012,Kawai2020,Yonezawa2016,Tao2018,Matano2016}, Sr$_x$Bi$_2$Se$_3$~\cite{Shruti2015,Liu2015,Kuntsevich2018,Kuntsevich2019,Pan2016,Neha2019,Bannikov2021}, Nb$_x$Bi$_2$Se$_3$~\cite{Qiu2015,Kurter2018,Asaba2017,Das2020} reveal in them a superconductivity below critical temperature $T_c \sim 3$K. The breaking of the rotational symmetry in these superconductors has been observed experimentally ~\cite{Yonezawa2016,Asaba2017,Pan2016,Kuntsevich2018,Tao2018} as well as the spin triplet character of the Cooper's pairing~\cite{Matano2016}. Such properties are best described if we assume that the superconducting order parameter is a two-component vector with $E_u$ representation, which is usually called nematic superconducting state~\cite{Fu2010,Fu2014,Venderbos2016,Yonezawa2018}. 

The nematic superconductivity generated a great interest due to its unusual properties such as existence of Majorana Kramer's pairs~\cite{Wu2017,Akzyanov2021a}, vestigial nematic order~\cite{Hecker2018}, surface Andreev bound states~\cite{Hao2017}, unconventional collective modes~\cite{Uematsu2019}, spontaneous strain~\cite{Akzyanov2020_2}, unusual magnetic response~\cite{Khokhlov2021a}, and anisotropic quasiparticle interference~\cite{Chen2018, Khokhlov2021}. A major of these effects are related to the vector nature of the order parameter. Orientation of the order parameter vector is often referred to as a nematicity direction and it controls the anisotropy axis of the system. Naturally, we expect an interesting Josephson physics in such superconductors, which depends significantly on the mutual orientations of the nematicity axis and the junction plane. 

In this work, we perform a study of the Josephson physics in the nematic topological superconductors of the type ABi$_2$Se$_3$ based on the phenomenological Ginzburg-Landau (GL) approach~\cite{Fu2014}. In Section~\ref{GL_equations}, we derive the GL equations for such superconductors. We get that the Meissner kernel that shows the response of the supercurrent to the vector potential has off-diagonal components. In Section~\ref{Current_phase}, we consider a superconductor-insulator-superconductor (SIS) Josephson junction and derive current-phase relations for different orientations of the junction plane and crystallographic axes. We show that the anomalous Josephson Hall effect (that is, the Josephson current along the junction) can be observed in the system for a certain orientation of the contact plane to the crystal axes. In Section~\ref{FP_eqs}, we analyze the electromagnetic properties of the junction and derive the dependence of the maximum Josephson current through the junction on the magnetic field. In Section~\ref{Discussion}, we discuss the obtained results.

\section{Ginzburg-Landau equations}\label{GL_equations}

We start with the GL free energy $F$ for the vector superconducting order parameter $\vec{\eta}=(\eta_1,\eta_2)$
\begin{equation}\label{GL1}
F=F_0+F_D+F_H, 
\end{equation}
where $F_0$ is a ``homogeneous" contribution, $F_D$ is the gradient term, and $F_H$ is a contribution due to the electromagnetic field. Following seminal paper by Fu, we write down $F_0$ in the form \cite{Fu2014} 
\begin{eqnarray}\label{F_0}
F_0&=&a(|\eta_1|^2+|\eta_2|^2)+B_1(|\eta_1|^2+|\eta_2|^2)^2\\
\nonumber
&+&B_2|\eta_1^*\eta_2-\eta_1\eta_2^*|^2,
\end{eqnarray}
where $a\propto T/T_c-1<0$ and $B_{1,2}$ are the GL coefficients, $T_c$ is the critical temperature. The gradient term for the doped topological insulator of the type ABi$_2$Se$_3$ can be chosen as~\cite{Fu2014,Zyuzin2017} 
\begin{widetext}
\begin{eqnarray}\label{F_D}
\nonumber
F_D&=&J_1(D_i\eta_{\alpha})^*D_i\eta_{\alpha} + J_3(D_z\eta_{\alpha})^*D_z\eta_{\alpha}+ J_4[(D_x\eta_{1})^*D_x\eta_{1}-(D_y\eta_{1})^*D_y\eta_{1})+(D_y\eta_{2})^*D_y\eta_{2}
\\
&-&(D_x\eta_{2})^*D_x\eta_{2}+(D_x\eta_{1})^*D_y\eta_{2}+(D_x\eta_{2})^*D_y\eta_{1}+(D_y\eta_{1})^*D_x\eta_{2}+(D_y\eta_{2})^*D_x\eta_{1}].
\end{eqnarray}
\end{widetext}
Here $J_{n}$ are corresponding GL coefficients, $D_j=-i\hbar\nabla_j-2eA_j/c$, summation over repeating indices $i=x,y$ and $\alpha=1,2$ is assumed, and $\mathbf{A}=(A_x,A_y,A_Z)$ is the vector-potential. The magnetic part of the free energy reads
\begin{equation}\label{F_H}
F_H=\frac{(\textrm{curl} {\bf A})^2}{8\pi}-\frac{\textrm{curl} {\bf A}\cdot {\bf H_0}}{4\pi},
\end{equation}
where $\mathbf{H_0}$ is the applied field.

\subsection{First GL equations}

The first GL equations can be obtained from the variation $\delta F/\delta \eta^*_i=0$. This was done in many papers (see, e.g., Ref.~\cite{Venderbos2016}). We present here the result for the sake of completeness
\begin{widetext}
\begin{eqnarray}\label{GL_I}
\left\{a\sigma_0+2B_1\eta^2\sigma_0+2B_2 \textrm{Im}\,(\eta^*_1\eta_2)\sigma_y 
+
(J_1D_x^2+J_1D_y^2+J_3D_z^2)\sigma_0+J_4 (D_x^2-D_y^2)\sigma_z+J_4[D_x,D_y] \sigma_x)\right\}\vec{\eta}=0.
\end{eqnarray}
\end{widetext}
Here $\sigma_j$ ($j=0,x,y,z$) are the Pauli matrices that act in $\vec{\eta}=(\eta_1,\eta_2)$ space, $\eta^2=|\eta_1|^2+|\eta_2|^2$, and $[D_x,D_y]=D_xD_y-D_yD_x$. Boundary conditions for the first GL equations at the superconductor-vacuum interface read
\begin{eqnarray}\label{eta1K}
J_1n_iD_i\eta_1+J_3n_zD_z\eta_1+\\ \nonumber J_4\left[n_xD_x\eta_1-n_yD_y\eta_1+n_xD_y\eta_2+n_yD_x\eta_2\right]=0,\\
\label{eta2K}
J_1n_iD_i\eta_2+J_3n_zD_z\eta_2+ \\ \nonumber J_4\left[n_yD_y\eta_2-n_xD_x\eta_2+n_xD_y\eta_1+n_yD_x\eta_1\right]=0,
\end{eqnarray}
where $n_i=n_{x,y}$ and $n_z$ are corresponding components of the external normal to the sample surface.

\subsection{Second GL equations}

To derive the second GL equations, we perform variation of the GL functional with respect to the components of the vector potential. Here we present the results in the form convenient for calculation of the Josephson current. We start with variation of $F_H$ and introduce, as usual, the current components 
\begin{eqnarray}\label{current_1}
\delta_{\mathbf A}F_H=\frac{1}{c}\delta{\mathbf A}\mathbf{j}_s, \qquad
{\mathbf j}_s=\frac{c}{4\pi}\textrm{curl}\, \textrm{curl} {\mathbf A}.
\end{eqnarray}
In so doing, we obtain the second GL equations from the condition $\delta_{\bf A}F=0$ in the form 
\begin{widetext}
\begin{eqnarray}\label{eqII}
j_{sz}&=&-2e\hbar J_3\left[F_{11}^z+F_{22}^z+\frac{4\pi}{\Phi_0} A_z(|\eta_1|^2+|\eta_2|^2)\right], \\
\nonumber
j_{sx}&=&-2e\hbar\left\{\!(J_1\!+\!J_4)\!\left(F_{11}^x\!+\!\frac{4\pi}{\Phi_0} A_x\eta_{1}^*\eta_{1}\right)\!+\!
(J_1\!-\!J_4)\!\left(F_{22}^x\!+\!\frac{4\pi}{\Phi_0} A_x\eta_{2}^*\eta_{2}\right)\!+\!
J_4\left[F_{12}^y\!+\!F_{21}^y\!+\!\frac{4\pi}{\Phi_0} A_y(\eta_{1}\eta_{2}^*\!+\!\eta_{1}^*\eta_{2})\right]
\!\right\}, \\
\nonumber
j_{sy}&=&-2e\hbar\left\{\!(J_1\!-\!J_4)\!\left(F_{11}^y\!+\!\frac{4\pi}{\Phi_0} A_y\eta_{1}^*\eta_{1}\right)\!+\!
(J_1\!+\!J_4)\!\left(F_{22}^y\!+\!\frac{4\pi}{\Phi_0} A_y\eta_{2}^*\eta_{2}\right)\!+\!
J_4\left[F_{12}^x\!+\!F_{21}^x\!+\!\frac{4\pi}{\Phi_0} A_x(\eta_{1}\eta_{2}^*\!+\!\eta_{1}^*\eta_{2})\right]
\!\right\}.
\end{eqnarray}
Here $F_{\alpha\beta}^j=i\eta_{\alpha}^*\nabla_j\eta_{\beta}-i\eta_{\beta}\nabla_j\eta_{\alpha}^*$ and $\Phi_0=\pi\hbar c/e$ is the magnetic flux quantum. 
We introduce phases of the vector order parameter as
$\eta_1=|\eta_1|e^{i\varphi_1}$, $\eta_2=|\eta_2|e^{i\varphi_2}$, $\varphi=(\varphi_1+\varphi_2)/2$, and $\delta=(\varphi_1-\varphi_2)/2$. We also express the order parameter components through a nematicity angle $\alpha$ as $|\eta_1|=\eta\cos{\alpha}$ and $|\eta_2|=\eta\sin{\alpha}$, $\alpha \in [0,\pi/2]$. 
In these notations we have
\begin{eqnarray*}
F_{11}^{j}&=&-2\eta^2\cos^2{\alpha} \nabla_j \varphi_1,\quad F_{22}^{j}=-2\eta^2\sin^2{\alpha} \nabla_j \varphi_2,\quad \\
F_{12}^{j}&+&F_{21}^{j}=-2\eta^2\left(\sin{2\alpha}\cos2\delta\nabla_j\varphi-\sin2\delta\nabla_j\alpha\right). 
\end{eqnarray*}
As a result, Eqs.~\eqref{eqII} reads
\begin{eqnarray}\label{jsz_alpha}
\!\!\!\!\!\!\!\!\!j_{sz}&=&4e\hbar J_3\eta^2\left(\nabla_z \varphi+\cos{2\alpha} \nabla_z \delta-\frac{2\pi}{\Phi_0}A_z\right),\\
\label{jsx_alpha}
\!\!\!\!\!\!\!\!\!j_{sx}\!&=&\!4e\hbar\eta^2\!\left\{\!(J_1\!+\!J_4\cos{2\alpha})\!\left(\!\nabla_x\varphi\!-\!\frac{2\pi}{\Phi_0}A_x\!\right)
\!+\!(J_1\cos{2\alpha}\!+\!J_4)\nabla_x\delta\!+\!J_4\!\left[\sin{2\alpha}
\cos{2\delta}\!\left(\!\nabla_y\varphi\!-\!\frac{2\pi}{\Phi_0}A_y\!\right)\!-\!\sin{2\delta}\nabla_y\alpha\!\right]\!\right\},\\
\label{jsy_alpha}
\!\!\!\!\!\!\!\!\!j_{sy}\!&=&\!4e\hbar\eta^2\!\left\{\!(J_1\!-\!J_4\cos\!{2\alpha})\!\left(\!\nabla_y\varphi\!-\!\frac{2\pi}{\Phi_0}A_y\!\right)
\!+\!(J_1\cos{2\alpha}\!-\!J_4)\nabla_y\delta\!+\!J_4\!\left[\sin{2\alpha}
\cos{2\delta}\!\left(\!\nabla_x\varphi\!-\!\frac{2\pi}{\Phi_0}A_x\!\right)\!-\!\sin{2\delta}\nabla_x\alpha\!\right]\!\right\}.
\end{eqnarray}
These rather cumbersome equations can be rewritten in a compact form
\begin{eqnarray}\label{GL_2_final}
j_{si}=4e\hbar\eta^2\left[\nu_i\left(\nabla_i\varphi-\frac{2\pi}{\Phi_0}A_i\right)\!+\!\gamma_i\nabla_i\delta+J_4\sin 2\alpha \cos 2\delta\left(\nabla_{\bar{i}}\varphi-\frac{2\pi}{\Phi_0}A_{\bar{i}}\right)-J_4\sin{2\delta\nabla_{\bar i}}\alpha\right].
\end{eqnarray}
\end{widetext}
Here $i=(x,y,z)$, $\bar{i}=(y,x,0)$, which mean that two last terms are absent in the formula for $j_{sz}$, $(\nu_x,\nu_y,\nu_z)=(J_1+J_4\cos 2\alpha,J_1-J_4\cos2\alpha,J_3)$, and $(\gamma_x,\gamma_y,\gamma_z)=(J_1\cos2\alpha+J_4,J_1\cos2\alpha-J_4,J_3\cos2\alpha)$. 

%Equations~\eqref{GL_2_final} must be gauge-invariant~\cite{Nambu1960}.
 Making a gradient transformation ${\bf A}\rightarrow {\bf A}+\Phi_0\nabla f/2\pi$ we restore the same Eq.~\ref{GL_2_final} after substitution $\varphi\rightarrow \varphi+f$. This means that $\varphi_i$ transform as $\varphi_i\rightarrow \varphi_i+f/2$. Note also, that the phase difference $\delta$ in the considered topological superconductors is a function of its internal properties and applied strain or magnetic field~\cite{Fu2014,Venderbos2016,Akzyanov2020_2}. For example, if applied strain and/or magnetic field are absent we have a pure nematic phase with $\delta=0$ when $B_2>0$ and a pure chiral phase with $\delta=\pi/2$ when $B_2<0$. The applied strain and/or magnetic field can transform the pure phase to some intermediate state with $0<\delta<\pi/2$~\cite{Akzyanov2020_2}. 

From the obtained formulas it follows, in particular, that the supercurrent along the $z$ direction is generated by the $z$ component of the vector potential and/or the phase gradient along the $z$ axis. However, in contrast to ``usual'' superconductors, the supercurrents along $x$ and $y$ directions are generated by both $A_x$ and $A_y$ components of the vector potential and $x$- and $y$-components of the phase gradient, as well. 

If we neglect variations of $\eta_{1,2}$ in the bulk of the superconductor we obtain the London equation for the nematic superconductor. In such a limit, it is convenient to rewrite Eq.~\eqref{GL_2_final} in the matrix form
\begin{equation}\label{London_current}
\mathbf{j}_s=-\frac{c}{4\pi}\hat{K}\mathbf{A}, \quad \hat{K}=\frac{32\pi e^2\eta^2}{c^2}\begin{pmatrix}
\nu_x & \bar{J}_4& 0\\
\bar{J}_4& \nu_y& 0\\
0& 0& \nu_z
\end{pmatrix}, 
\end{equation}
where $\bar{J}_4=J_4\sin 2\alpha\cos{2\delta}$. 
As we see, Meissner kernel $K_{\alpha \beta}$ has off-diagonal components $K_{xy}$ which is unusual. The $K_{xy}$ terms in the Meissner kernel are inherent for superconductors anisotropic in the $xy$-plane. The considered materials are crystallographically isotropic in this plane. The anisotropy arises in the superconducting state due to the vector nature of the order parameter. Note, that the existence of the off-diagonal components of the Meissner kernel has been missed in previous microscopic calculations~\cite{Schmidt2020, Akzyanov2021b}.

Substituting Eq.~\eqref{London_current} in the corresponding Maxwell equation, we obtain the London equation for the nematic superconductor
\begin{equation}\label{London_Eq}
\mathbf{H}+\textrm{curl}\left(\hat{K}^{-1}\textrm{curl}\right)\,\mathbf{H}=0. 
\end{equation}
Since $\hat{K}^{-1}$ has off-diagonal components, we have a mixing between $H_x$ and $H_y$ components of the magnetic field.

\section{Current-phase relations}\label{Current_phase}

In this section, we obtain relations between the Josephson current through a SIS junction and the phases of the order parameter components when the external electromagnetic field is absent. The junction consists of two superconductors with the nematic pairing. Superconducting correlations are induced by the proximity effect inside the junction $-d/2<z<d/2$. We assume that these correlations are weak~\cite{Buzdin2008} and we can keep only quadratic in $\eta_i \eta_j$ terms neglecting terms of the fourth order that leads to the linearized GL equations that we obtain from ~\eqref{GL1}. Following a standard approach, we assume that the GL coefficient in the junction $a=a_n$ is positive. 

\subsection{Junction transverse to z direction}

\begin{figure}[b]
\includegraphics[width=1.0\linewidth]{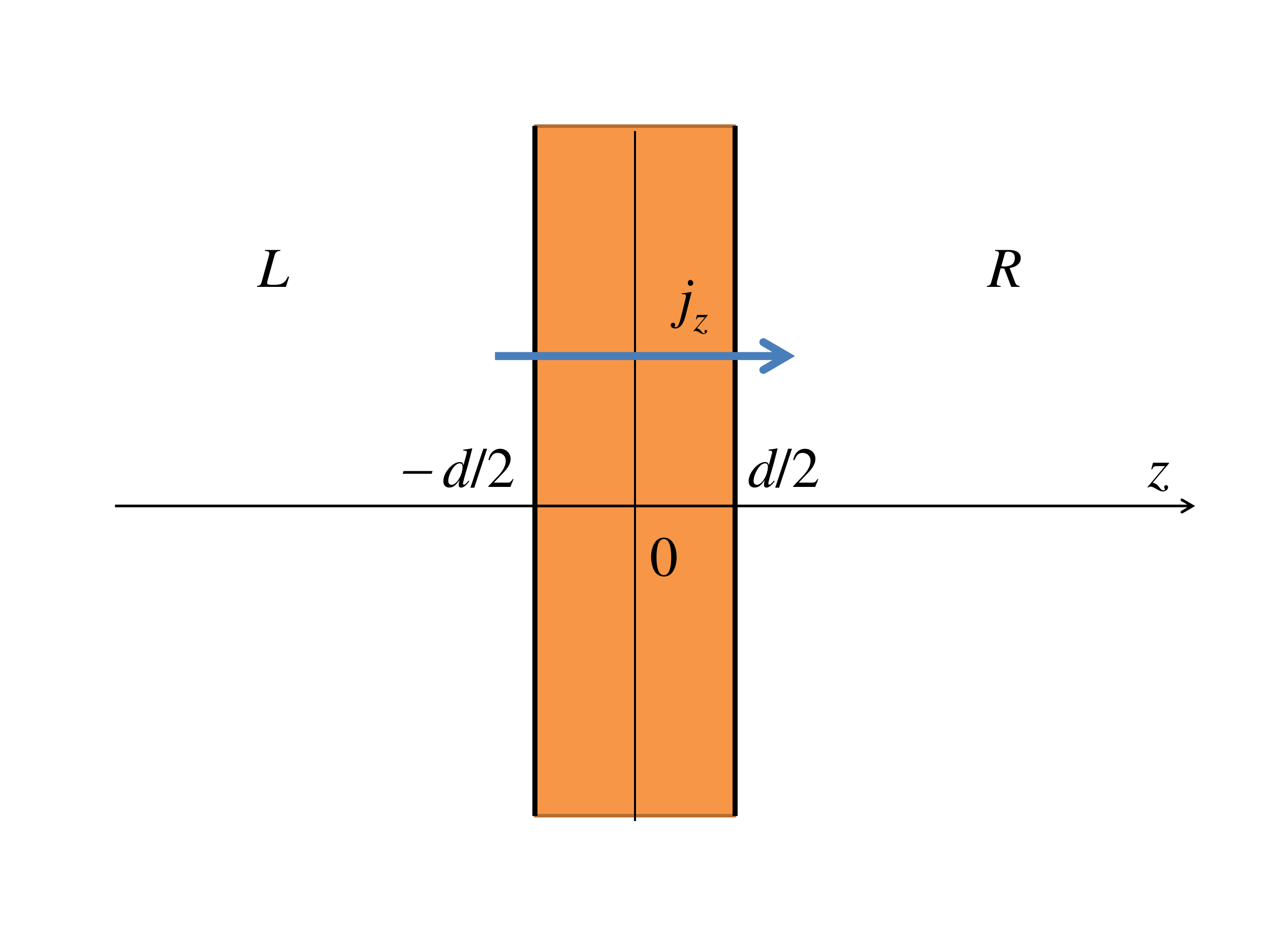}
\caption{Junction transverse to $z$ direction.}
%%%%%%%%%%%%%%%%%%%%%%%%%%%%%%%%%%%%%%%%%%%%%%%%%%%%%%%%%%
\label{J_z}
%%%%%%%%%%%%%%%%%%%%%%%%%%%%%%%%%%%%%%%%%%%%%%%%%%%%%%%%%%%
\end{figure}

Consider a junction transverse to $z$ direction, which we will call further $z$-junction Fig.~\ref{J_z}. Inside the junction $-d/2<z<d/2$ the linearized GL equation~\eqref{GL_I} reads as
\begin{equation}
\nabla_z^2 \vec{\eta}= \frac{a_N}{\hbar^2J_3}\vec{\eta}.
\end{equation}
We will use subscript $L$ for the parameters at the left ($z<0$) of the junction and $R$ for the parameters at the right ($z>0$). 
We set the boundary conditions $\eta_j(+d/2)=\eta_{jR}$ ans $\eta_j(-d/2)=\eta_{jL}$ and obtain
\begin{eqnarray}\label{eta_z}
\!\!\!\eta_j\!=\!\frac{(\eta_{jR}\!+\!\eta_{jL})\cosh \kappa_z z}{2\cosh (\kappa_z d/2)}\!+\!\frac{(\eta_{jR}\!-\!\eta_{jL})\sinh \kappa_z z}{2\sinh (\kappa_z d/2)},
\end{eqnarray}
where $\kappa_z=\sqrt{a_N/\hbar^2J_3}$. We substitute this formula into Eq.~\eqref{GL_2_final} for the supercurrent along $z$ axis and derive
\begin{eqnarray}\label{j_z_theta12}
\!\!\!\!\!\!j_z\!=\!j_{cz}\left[\cos{\alpha_R}\cos{\alpha_L}\sin{\theta_1\!}+\!\sin{\alpha_R}\sin{\alpha_L}\sin{\theta_2}\right], 
\end{eqnarray}
where $\theta_j=\varphi_{jR}-\varphi_{jL}$ and 
\begin{equation}\label{j_cz}
j_{cz}=\frac{4e\hbar J_3\kappa_z\eta_R\eta_L}{\sinh \kappa_z d}. 
\end{equation}
We see that the Josephson current through the junction is a sum of contributions from two components of the order parameter. This relation differs from a general current-phase relation for usual two-band superconductors~\cite{Sasaki2020} since the phases of the order parameter components in the nematic superconductors are not independent~\cite{Fu2014}. It is reasonable to express the Josephson current in terms of the sum and difference of the phases of the order parameter components:
\begin{eqnarray}\label{j_z_theta}
&&j_z=j_{cz}\times\\
\nonumber
&&\left[\cos(\alpha_R\!-\!\alpha_L)\sin{\theta}\cos{\phi}\!+\!\cos(\alpha_R\!+\!\alpha_L)\cos{\theta}\sin{\phi}\right], 
\end{eqnarray}
where $\theta=\varphi_{R}-\varphi_{L}$ and $\phi=\delta_{R}-\delta_{L}$. We can tune the value of the Josephson current varying the nematicity direction. For example, in the case of the Josephson junction between two purely nematic superconductors $\delta_L=\delta_R=0$ expression for the Josephson current takes a simple form $j_z=j_{cz} \cos(\alpha_R\!-\!\alpha_L)\sin \theta$. We see that if the vector order parameters at different sides of the contact are orthogonal, $\alpha_R-\alpha_L=\pi/2$ there is no current through the junction, $j_z=0$. In other terms, $j_z=0$ if $\vec{\eta}_R=\eta_R(1,0)$ and $\vec{\eta}_L=\eta_L(0,1)$ or vice versa $\vec{\eta}_R=\eta_R(0,1)$ and $\vec{\eta}_L=\eta_L(1,0)$. 

As usual for the Josephson effect, we assume that the density of superconducting current in the bulk is much larger than the Josephson critical current density~~\cite{barone1982, Tinkham2004, Likharev1979}. That is, the London penetration depth $\lambda$ is much larger than a Josephson penetration depth $\lambda_J$. To estimate $\lambda_J$ we need equations for the junction in the magnetic field obtained below in Section~\ref{FP_eqs}. Here we only note that the condition $\lambda\ll\lambda_J$ requires $\kappa d\gg 1$, that is, the contact is far from a superconducting transition, and the value $\eta(z=0)$ is much smaller than the equilibrium order parameter in the bulk, $\eta(|z|>d/2)$.

\subsection{Junction transverse to x direction}

\begin{figure}[b]
\includegraphics[width=1.0\linewidth]{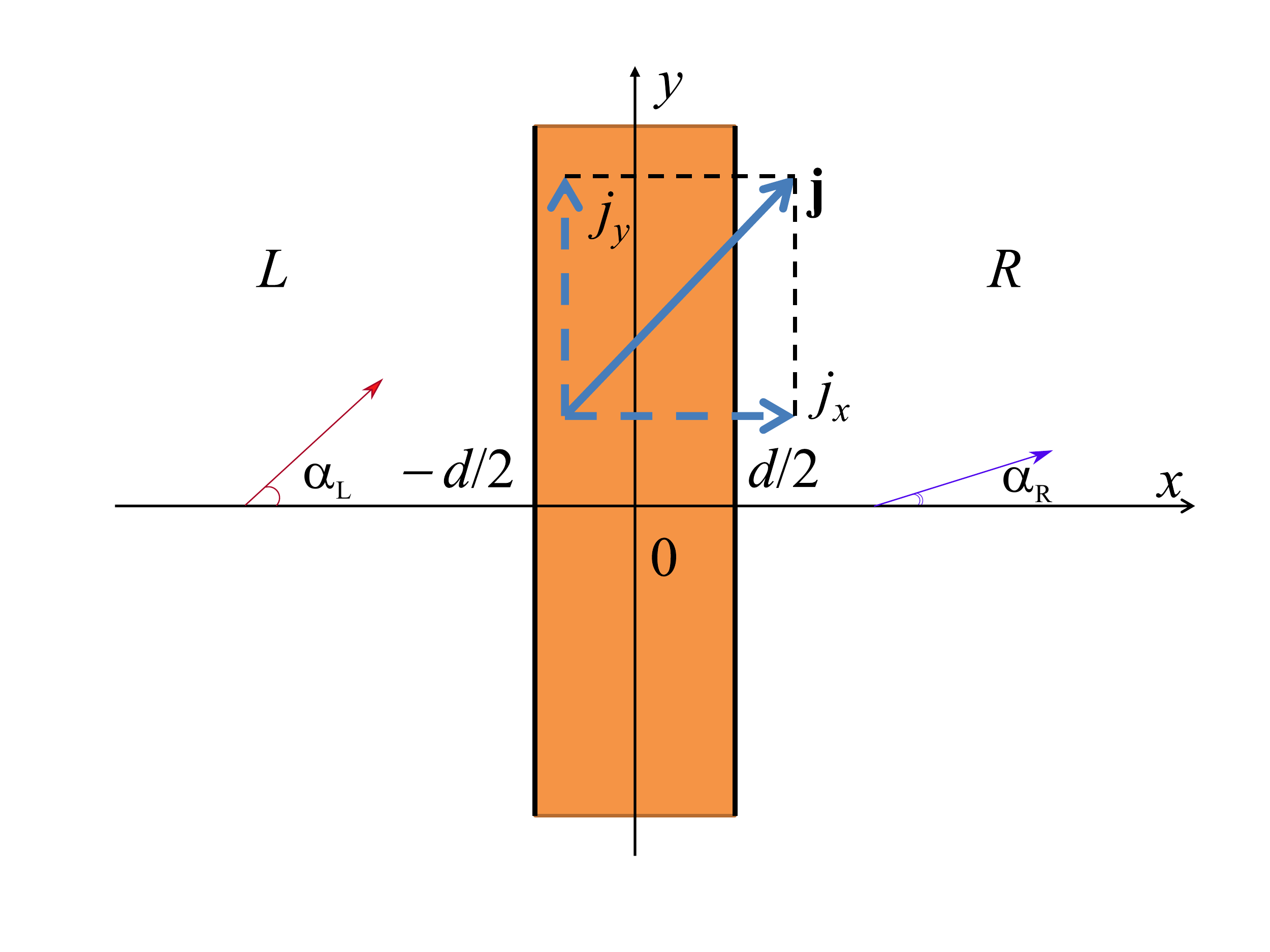}
\caption{Junction transverse to $x$ axis. Nematicity angles $\alpha_L$ and $\alpha_R$ show direction of the vector of the order parameter, $\delta_{L(R)}=0$.}
%%%%%%%%%%%%%%%%%%%%%%%%%%%%%%%%%%%%%%%%%%%%%%%%%%%%%%%%%%
\label{J_x}
%%%%%%%%%\alpha%%%%%%%%%%%%%%%%%%%%%%%%%%%%%%%%%%%%%%%%%%%%%%%%,
\end{figure}

Here we consider a junction transverse to the $x$ axis, $x$-junction (see Fig.~\ref{J_x}). The case of $y$-junction is a similar. The linearized first GL equation~\eqref{GL_I} in the junction $-d/2<x<d/2$ is
\begin{equation}\label{GL_I_x}
\nabla_x^2 \vec{\eta}=\frac{(J_1{\color{blue}-}J_4\sigma_z)\,a_N}{\hbar^2(J_1^2-J_4^2)}\vec{\eta}. 
\end{equation}
We set boundary conditions $\eta_j(+d/2)=\eta_{jR}$ and $\eta_j(-d/2)=\eta_{jL}$ and derive
\begin{equation}\label{eta_x}
\!\!\!\eta_j\!=\!\frac{(\eta_{jR}\!+\!\eta_{jL})\cosh \kappa_j x}{2\cosh (\kappa_j d/2)}\!+\!\frac{(\eta_{jR}\!-\!\eta_{jL})\sinh \kappa_j x}{2\sinh (\kappa_j d/2)}, 
\end{equation}
where $\kappa_1=\sqrt{a_N/\hbar(J_1+J_4)}$ and $\kappa_2=\sqrt{a_N/\hbar(J_1-J_4)}$. 
We substitute Eq.~\eqref{eta_x} in the formula for the components of the supercurrent, Eqs.~\eqref{GL_2_final}. We observe that in the considered geometry the Josephson current through the junction has two components, usual current through the junction $j_{x}$ and transverse or Hall current along it, $j_{y}$. For the current through the contact we obtain
\begin{eqnarray}\label{j_x_theta12}
j_x&=&j_{c1}\cos{\alpha_R}\!\cos{\alpha_L}\!\sin{\theta_1}\!\!+\!j_{c2}\sin{\alpha_R}\sin{\alpha_L}\!\sin{\theta_2}, \\
\nonumber
j_{c1}\!&=&\!\frac{4e\hbar(J_1\!+\!J_4)\kappa_1\eta_L\eta_R}{\sinh \kappa_1 d}, \,\,\, j_{c2}\!=\!\frac{4e\hbar(J_1\!-\!J_4)\kappa_2\eta_L\eta_R}{\sinh \kappa_2 d}.
\end{eqnarray}
As in the case of the $z$ junction, $j_x$ consists of two terms, which corresponds to two components of the order parameter, but in contrast to the previous case, the critical values of the current for $\eta_1$ and $\eta_2$ are different. Similar to Eq.~\eqref{j_z_theta}, we can rewrite the latter formula in terms of $\theta$ and $\phi$
\begin{eqnarray}\label{j_x_theta}
\nonumber
&&j_x=\left(j_{c1}\cos{\alpha_R}\!\cos{\alpha_L}\!+\!j_{c2}\sin{\alpha_R}\!\sin{\alpha_L}\right)\sin{\theta}\cos{\phi} \\
&&+\left(j_{c1}\cos{\alpha_R}\!\cos{\alpha_L}\!-\!j_{c2}\sin{\alpha_R}\!\sin{\alpha_L}\right)\cos{\theta}\sin{\phi}. 
\end{eqnarray}
As in the case of the $z$-junction, we can tune the current value and even ``turn off'' the contact (that is, to put $j_x=0$) by a proper choice of the nematicity direction. As an example, consider a junction formed by two purely nematic superconductors, $\delta_R=\delta_L=\phi=0$. In this case, $j_x=0$ if the nematicity direction in right and left sides of the junction are perpendicular, just similar to the case of the $z$-junction. 

\begin{figure}[b]
\includegraphics[width=1\linewidth]{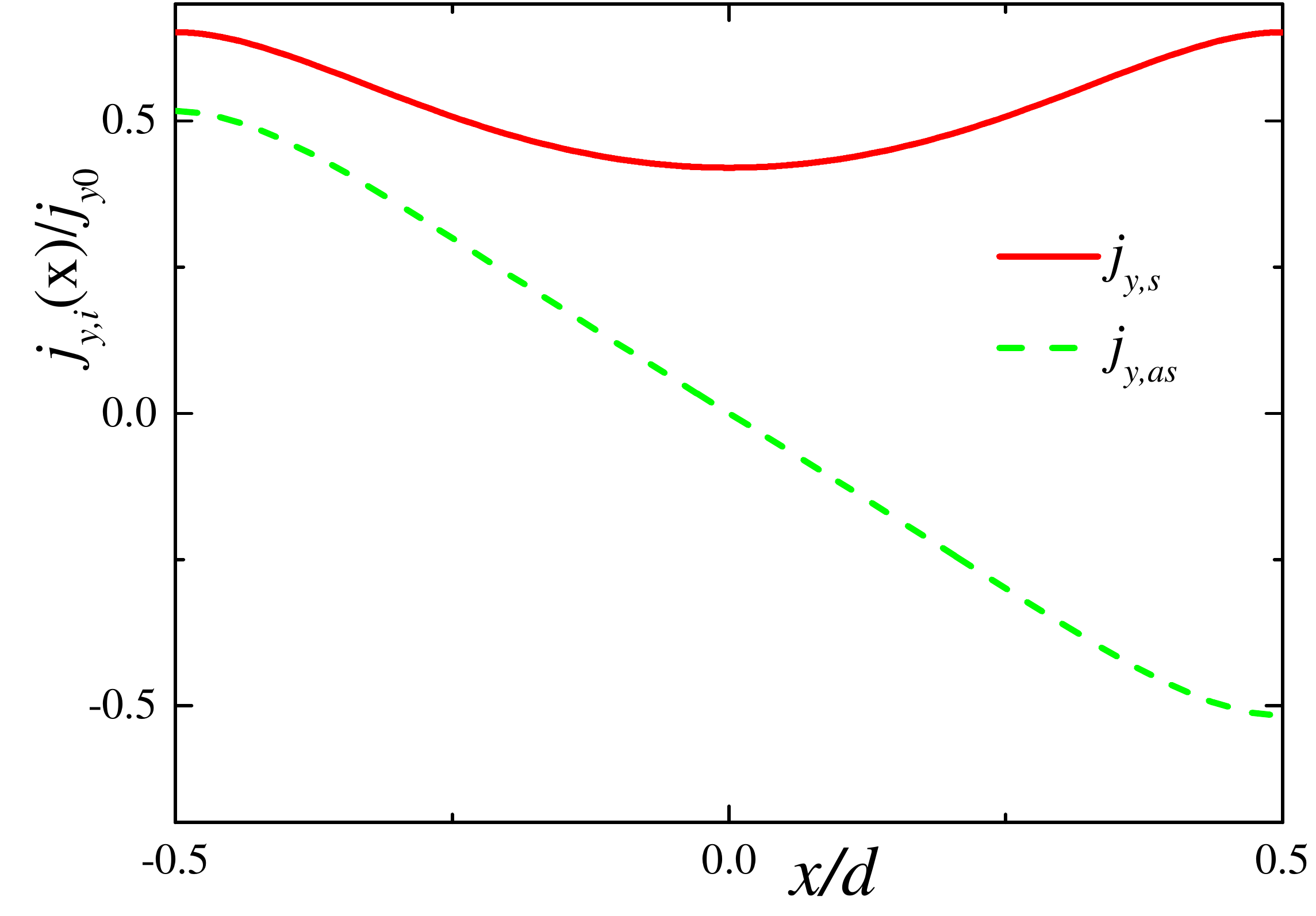}
\caption{Dependence of the symmetric $j_{y,s}$ (red line) and antisymmetric $j_{y,as}$ (green dashed line) components of the transverse current $j_y(x)/j_{y0}$ on the coordinate $x$ for $J_4/J_1=3/4$, $k_1d = 5$. Here $j_{y0}=4e\hbar J_4 $}
%%%%%%%%%%%%%%%%%%%%%%%%%%%%%%%%%%%%%%%%%%%%%%%%%%%%%%%%%%
\label{Figjyx}
%%%%%%%%%%%%%%%%%%%%%%%%%%%%%%%%%%%%%%%%%%%%%%%%%%%%%%%%%%%
\end{figure}

\begin{figure}[b]
\includegraphics[width=1\linewidth]{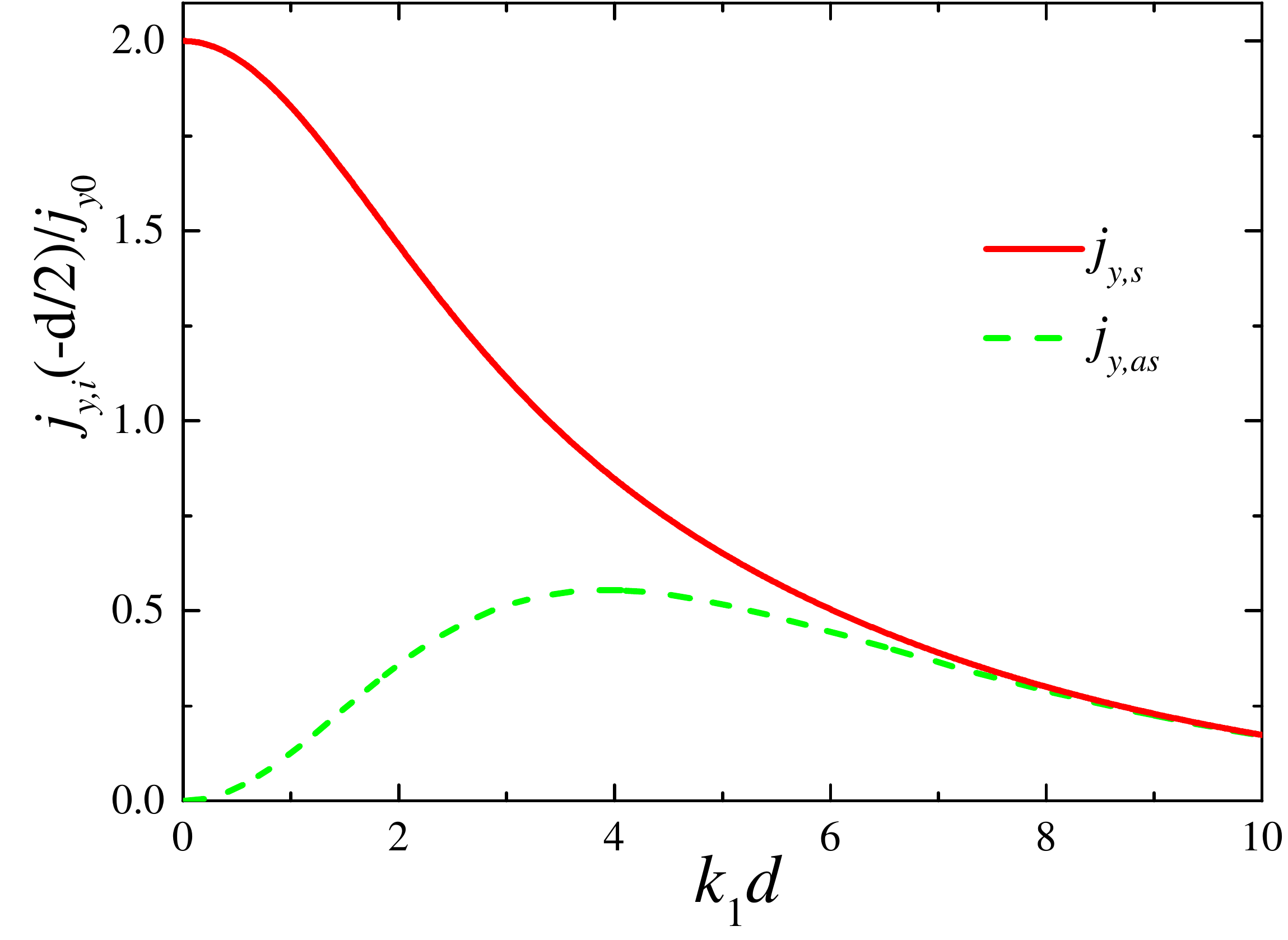}
\caption{Dependence of the symmetric $j_{y,s}$ (red line) and antisymmetric $j_{y,as}$ (green dashed line) components of the transverse current $j_y(x)/j_{y0}$ on the $k_1d$ for $J_4/J_1=3/4$ at the left side of the contact $x=-d/2$. Here $j_{y0}=4e\hbar J_4 $}
%%%%%%%%%%%%%%%%%%%%%%%%%%%%%%%%%%%%%%%%%%%%%%%%%%%%%%%%%%
\label{Figjkd}
%%%%%%%%%%%%%%%%%%%%%%%%%%%%%%%%%%%%%%%%%%%%%%%%%%%%%%%%%%%
\end{figure}

According to Eq.~\eqref{GL_2_final}, in addition to $j_x$, we have a Josephson Hall current $j_y$ flowing along the junction. Presence of such a transverse current is a direct consequence of the non-diagonal Meissner kernel and is an analog of the anomalous Josephson Hall effect~\cite{Wang2011,Vedyayev2013,Yokoyama2015,MatosAbiague2015,Dang2015,Mironov2017,Malshukov2019, Costa2020, Maistrenko2021}. Using Eqs.~\eqref{eqII} and~\eqref{eta_x} we obtain
\begin{eqnarray}\label{jy_x}
j_{y}=4e\hbar J_4 \textrm{Im}\left(\eta_1^*\nabla_x\eta_2+\eta_2^*\nabla_x\eta_1\right).
\end{eqnarray}
The current $j_{y}$ depends on $x$. In the case of two nematic superconductors, the function $j_y(x)$ can be presented as $j_y(x)=j_{y,s}(x)\sin(\alpha_R+\alpha_L)+j_{y,as}(x) \sin (\alpha_R-\alpha_L)$ where $j_{y,s}(x)=j_{y,s}(-x)$ is a symmetric component of the current and $j_{y,as}(x)=-j_{y,as}(-x)$ is antisymmetric. As we see, tuning of $\alpha_R$ and $\alpha_L$ can give us purely symmetric or antisymmetric current. In general, the transverse current $j_y$ can be of the same order or even larger (for $\delta >0$) than $j_x$. Moreover, $j_y$ can be non-zero when the current through the junction $j_x$ is zero. The coordinate dependence of the symmetric and antisymmetric parts of the current $j_y(x)$ is illustrated in Fig.~\ref{Figjyx}. The dependence of these components of the current on the junction thickness $d$ is shown in Fig.~\ref{Figjkd}. It is interesting to note that $j_{y,as}(d)$ first grows with $d$, attains a maximum, and finally decreases, while $j_{y,s}(d)$ decreases monotone from maximum at small $d$ to zero, when $\kappa_1d\gg 1$, as it is common for Josephson currents~\cite{barone1982}.

Consider the case of the junction between two nematic superconductors, $\delta_R=\delta_L=0$. To compare $j_y$ and $j_x$, we introduce an average value
\begin{equation}\label{average}
\bar{j}_y=\frac{1}{d}\int_{-d/2}^{d/2}j_y(x)dx 
\end{equation}
After a straightforward calculations we derive
\begin{equation}\label{Hall_current}
\bar{j}_y=j_{cy}\sin{\theta}, 
\end{equation}
where
\begin{align}
j_{cy}=\sin(\alpha_R+\alpha_L)\frac{8e\hbar\eta^2\sqrt{J_1^2-J_4^2}}{d\sinh{(\kappa_1d)}\sinh{(\kappa_2d)}}\\
\nonumber
\times(\cosh{\kappa_2d}-\cosh{\kappa_1d})
\end{align}
In the case of two nematic superconductors $\delta_R=\delta_L=0$, $j_y$ and $j_x$ are typically of the same order for $\kappa_i d \sim 1$ and $j_{cy}/j_{cx} \sim \tan{\alpha}/\kappa_1 d$ for $\kappa_i d \gg 1$. The value of the Hall current $\bar{j}_y$ is zero in the follows cases: (i) if $J_4=0$ (that is, $\kappa_1=\kappa_2$) and the anisotropy is absent in $(x,y)$ plane; (ii) if $\alpha_R+\alpha_L=0$ or $\alpha_R+\alpha_L=\pi$. 

\section{Junction in the magnetic field}\label{FP_eqs} 

In this section, we consider the junction in the magnetic field $\mathbf H$. As usual for Josephson physics, we assume that $\mathbf H$ lies in the contact plane. We consider here only the case of two purely nematics superconductors, $\delta_L=\delta_R=0$. Non-zero $\delta$ results in the existence of an additional magnetization in the superconductor, see Ref.~\cite{Akzyanov2020_2}, which, in general, should be included in the Maxwell equations. Also, finite $\delta$ leads to the presence of the antisymmetric component in the current along the junction, see the text after Eq.~\eqref{jy_x}. These effects deserve a separate detailed study in future work.

\subsection{z-junction}

First, we consider the junction, which lies in the plane $(x,y)$ transverse to z-axis. Since the problem has rotation symmetry in the $(x,y)$ plane we can choose the $y$ axis along the direction of the applied magnetic field, ${\mathbf H}_0=(0,H_0,0)$. The magnetic field in the junction, $z=0$, is ${\mathbf H}(x,y)=[0,H(x,y),0]$ and
\begin{equation}\label{Maxwell_junction}
\nabla_x H(x,y)=\frac{4\pi}{c} j_{z}, 
\end{equation}
where $j_{z}$ is the current through the junction.

First, we consider the case without Josephson currents. 
The magnetic field in the bulk of the superconductor, $\mathbf{H}(z)=[H_x(z),H_y(z),0]$ we can find using London equation~\eqref{London_Eq}. In this limit we have 
\begin{eqnarray*}
\nabla^2_z H_x&=&\frac{32\pi e^2\eta^2}{c^2}\left(\nu_yH_x-\bar{J}_4H_y\right),\\
\nabla^2_z H_y&=&\frac{32\pi e^2\eta^2}{c^2}\left(\nu_xH_y-\bar{J}_4H_x\right),
\end{eqnarray*}
and the dependence of $\mathbf{H}$ on $x$ is weak. The solution of these system decays when $z\rightarrow\pm\infty$ and $\mathbf{H}(z=0)=(0,H(x,y),0)$. In the considered case $\delta=0$ (purely nematic superconductor) we obtain inside the superconductor $|z|>d/2$ 
\begin{eqnarray}\label{penetration_z}
\nonumber
H_y&=&H(x,y)\!\left(\sin^2\!{\alpha_{L(R)}}\,e^{\frac{d/2-|z|}{\lambda_1}}+\cos^2\!{\alpha_{L(R)}}\,e^{\frac{d/2-|z|}{\lambda_1}}\right),\\
\nonumber
H_x&=&\frac{H(x,y)\sin{2\alpha_{L(R)}}\!\left(e^{\frac{d/2-|z|}{\lambda_2}}\!-\!e^{\frac{d/2-|z|}{\lambda_1}}\right)\textrm{sign}(z)}{2}, \\
\lambda_{1,2}&=&\frac{c}{4\sqrt{2\pi(J_1\pm J_4)}\,e\eta}. 
\end{eqnarray}
Thus, we obtain an estimate for the London penetration depths $\lambda_{1,2}$ and observe that the magnetic field in the $y$-direction in the junction generates the $x$-component of the field in the bulk. The latter property is a characteristic of anisotropic superconductors~\cite{Kogan1981}. In the considered geometry the anisotropy in the $(x,y)$ plane arises due to the vector nature of the order parameter.

The relation between the phase difference in the junction and magnetic field can be derived by different methods giving the same result~\cite{Ferrell1963, Kulik1972,barone1982, Likharev1979}. We outline briefly one of them.

From the Maxwell equation we have 
\begin{eqnarray}\label{Maxwell_jz}
j_x=-\frac{c}{4\pi}\nabla_zH_y, \quad j_y=\frac{c}{4\pi}\nabla_zH_x.
\end{eqnarray}
We integrate the components of the magnetic field across the junction and using Eqs.~\eqref{penetration_z}
obtain
\begin{equation}\label{definition_dz}
\int_{-\infty}^{+\infty}H_y(z)dz=Hd_1,\quad \int_{-\infty}^{+\infty}H_x(z)dz=Hd_2, 
\end{equation}
where 
\begin{eqnarray}\label{d1d2}
\nonumber
d_1&=&d\!+\!(\sin^2{\alpha_R}\!+\!\sin^2{\alpha_L})\lambda_1\!+\!(\cos^2{\alpha_R}\!+\!\cos^2{\alpha_L})\lambda_2,\\
d_2&=&d/2+(\sin{2\alpha_R}-\sin{2\alpha_L})(\lambda_2-\lambda_1)/2.
\end{eqnarray}
The first of these equations is a formal definition of the effective ``magnetic thickness'' of the junction $d_1$. We assume here for definiteness that $\sin{2\alpha_R}-\sin{2\alpha_L}>0$ since the choice of the right and left sides of the junction is arbitrary.  

We differentiate Eqs.~\eqref{Maxwell_jz} with respect to z having in mind that $\nabla_z\varphi\approx \theta\delta(z)$ if $\lambda_J\gg\lambda_{1,2}$ [here $\delta(z)$ is delta-function]. Then, we integrate the result with respect to $z$ from $z>-z_1$ to $z<z_1$, where $z_1\gg d_1/2$. Since $H_{x,y}(z)$ tends to zero with all its derivatives when $z\rightarrow\infty$, we come to relations
\begin{equation}\label{phase_A_z}
\nabla_i\theta=\frac{2\pi}{\Phi_0}\int_{-\infty}^{+\infty}\nabla_z A_i dz, 
\end{equation}
where $i=x,y$. Taking into account that $H_x=-\nabla_zA_y$, $H_y=\nabla_zA_x$, after integration we derive
\begin{align}\label{H_theta_z}
\nabla_x\theta=\frac{2\pi d_1}{\Phi_0}H(x,y), \quad
\nabla_y\theta=\frac{2\pi d_2}{\Phi_0}H(x,y).
\end{align}
Finally, using Eqs.~\eqref{j_z_theta} and~\eqref{Maxwell_junction} we obtain the equation for $\theta$ in the form 
\begin{align}\label{F_P_ZY}
\lambda_J^2\theta''_{xx}= \sin{\theta}, \\
\theta'_y=\frac{d_2}{d_1} \theta'_x
\end{align}
where we introduce the Josephson length in the usual form~\cite{barone1982}
\begin{equation}\label{lambda_J_z}
\lambda_J=\sqrt{\frac{c\Phi_0}{8\pi^2d_1j_{cz}\cos{(\alpha_R-\alpha_L)}}}. 
\end{equation}
Note, that for $\alpha_R=\alpha_L$ we have $d_2=0$ and $\theta'_y=0$.

Equation~\eqref{F_P_ZY} is often called Ferrell-Prange equation~\cite{Ferrell1963}. The first of Eqs~\eqref{H_theta_z} relays $\theta$ and the magnetic field and also gives boundary condition for $\theta$ at the edges of the junction where $H=H_0$. The values $\alpha_{R,L}$ are constants, which depend on the superconducting state in the bulk. We can rewrite the applicability condition of our approach $\lambda_J\gg\lambda_{1,2}$ in the explicit form using Eqs.~\eqref{j_cz}, \eqref{penetration_z}, and~\eqref{lambda_J_z} 
\begin{equation}\label{applic_z}
\frac{(J_1\pm J_4)\sinh{\kappa_zd}}{J_3\kappa_zd_1}\gg 1. 
\end{equation}

\begin{figure}[h]
\includegraphics[width=1.0\linewidth]{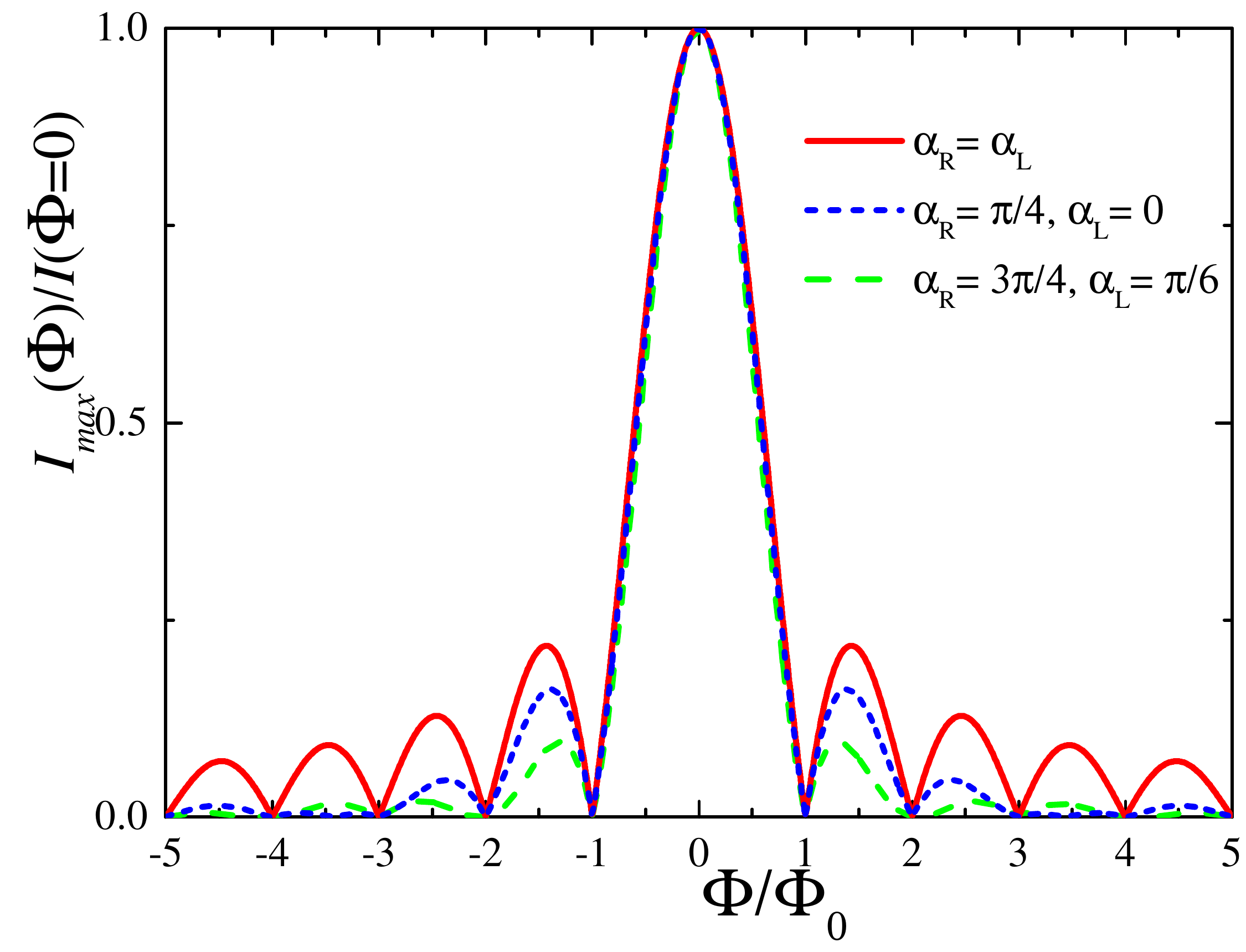}
\caption{Dependence of the normalized maximum critical current on the magnetic flux for different nematicity angles $\alpha_R$ and $\alpha_L$. We set $J_4/J_1=3/4$, $L_y=L_x$ and $d \ll \lambda_1$.}
%%%%%%%%%%%%%%%%%%%%%%%%%%%%%%%%%%%%%%%%%%%%%%%%%%%%%%%%%%
\label{fraun}
%%%%%%%%%%%%%%%%%%%%%%%%%%%%%%%%%%%%%%%%%%%%%%%%%%%%%%%%%%%
\end{figure}

\begin{figure}[h]
\includegraphics[width=1.0\linewidth]{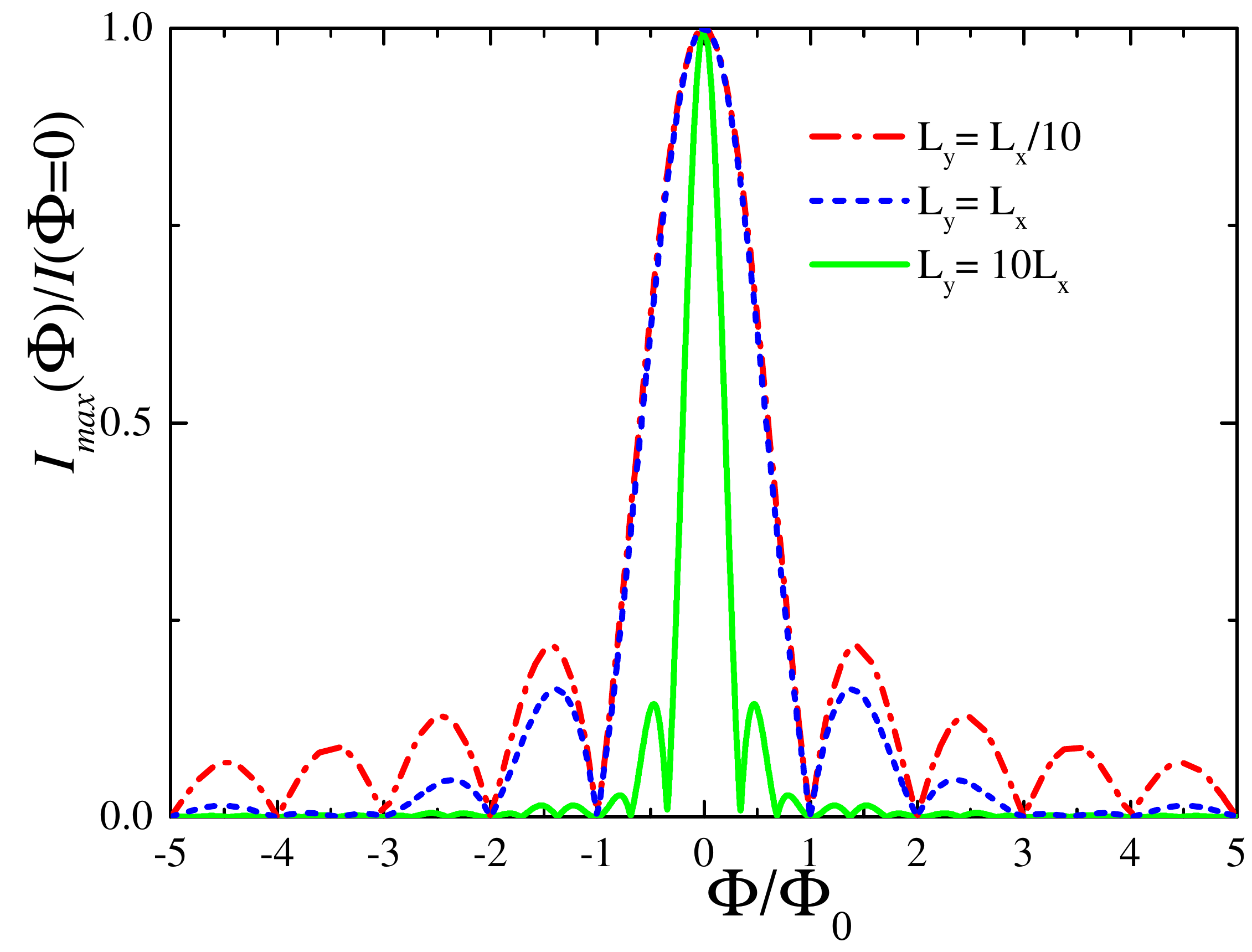}
\caption{Dependence of the normalized maximum critical current on the magnetic flux for different values of the aspect ratio of the contact $L_y/L_x$. We set $J_4/J_1=3/4$, $\alpha_R=\pi/4$, $\alpha_L=0$ and $d \ll \lambda_1$.}
%%%%%%%%%%%%%%%%%%%%%%%%%%%%%%%%%%%%%%%%%%%%%%%%%%%%%%%%%%
\label{fraun_ly}
%%%%%%%%%%%%%%%%%%%%%%%%%%%%%%%%%%%%%%%%%%%%%%%%%%%%%%%%%%%
\end{figure}

Now we derive a dependence of the maximum persistent current through the junction $I_{max}$ on the applied magnetic field $H_0$ or on the value of the captured in the junction magnetic flux $\Phi$~\cite{barone1982}. As usual, we consider a short contact in $x$ and $y$ directions $L_{x,y}\ll 2\lambda_J$ placed in a sufficiently large magnetic field, $H_0\gg \Phi_0/2\pi\lambda_Jd_1$, where $L_i$ is the junction length in $i$-th direction. Within these limits, the magnetic field in the junction is homogeneous and equals the applied field. We integrate the first of equations~\eqref{H_theta_z} taking into account that $H(x,y)=H_0$ and obtain
\begin{equation}\label{theta_H_z}
\theta=\frac{2\pi\Phi}{\Phi_0}\left(\frac{x}{L_x}+\frac{d_2}{d_1}\frac{y}{L_x}\right)+C, 
\end{equation}
where $\Phi=L_xd_1H_0$ is the magnetic flux captured in the junction and $C$ is a constant. 
We substitute the latter expression in Eq.~\eqref{j_z_theta} for the current and integrate it along the junction from $x=-L_x/2$ to $x=L_x/2$ and $y=-L_y/2$ to $y=L_y/2$. We obtain the value of the current through the contact $I$ in the form 
\begin{equation*}
\frac{I}{I_{cz}}=\cos(\alpha_L-\alpha_R)\sin{C}\frac{\sin{(\pi\Phi/\Phi_0)}}{\pi\Phi/\Phi_0}\frac{\sin{(\kappa \pi\Phi/\Phi_0)}}{\kappa\pi\Phi/\Phi_0}, 
\end{equation*}
where $I_{cz}=j_{cz}dL_x$, $\kappa = d_2L_y/d_1L_x$. Thus, the maximum of $I$ is 
\begin{equation}\label{Imax}
\!\!\!\frac{I_{max}}{I_{cz}}\!=\!\left|\cos(\alpha_L\!-\!\alpha_R)\frac{\sin{(\pi\Phi/\Phi_0)}}{\pi\Phi/\Phi_0}\frac{\sin{(\kappa\pi\Phi/\Phi_0)}}{\kappa\pi\Phi/\Phi_0}\right|, 
\end{equation}
%%%%%%%%%%%%%%%%%%%%%%%%%%%%%%%%%%%%%%%%%%%%%%%%%
The dependence $I_{max}(\Phi)$ is shown in Fig.~\ref{fraun}. It has a typical Fraunhofer-like pattern, but details of this pattern  differ from that for a usual $s$-wave superconductor (that corresponds to $d_2=0$ for the considered problem). Mathematically, this difference occurs due to dependence of the phase $\theta$ on two coordinates, $\theta=\theta(x,y)$. One of the prominent feature of the observed picture is that the maximal critical current decays as $I_{max} \propto 1/H^2$ instead of  $I_{max} \propto 1/H$ for a the $s$-wave superconductors. Period of the current $I_{max}(\Phi)$ oscillations depends on the parameters of the system, in particular, on the aspect ratio of the junction $L_y/L_x$. The latter feature is illustrated in Fig.~\ref{fraun_ly}. The variation of the oscillationa period and $1/H^2$ decay of the maximum current occurs due to additional factor $\sin (\kappa f)/\kappa f$ in Eq.~\ref{Imax} where $f=\pi\Phi/\Phi_0$. If $L_y \ll L_x$, we get $\kappa \ll 1$ and $I_{max} \propto |\sin f/f|$, which corresponds to the case of the s-wave superconductor. If $L_y \gg L_x$ and $\kappa \gg 1$, then, the period of the oscillations becomes significantly smaller than $\Phi_0$ when the applied magnetic field is low. Note, that we can control behavior of $I_{max}(\Phi)$ rotating the applied field in the junction plane if values $L_x$ and $L_y$ are significantly different. 


\subsection{x-junction}

Here we consider a junction transverse to the $x$-axis and the magnetic field lying in the $(y,z)$ plane, $\mathbf{H}(x=0)=(0,H_y,H_z)$, where $H_{y,z}$ depend slowly on $y$ and $z$. Using London equation~\eqref{London_Eq} and following a similar procedure as for the $z$-junction, we derive for the magnetic field in the bulk of the superconductor
\begin{eqnarray}\label{H_x_junction}
\nonumber
H_y(x)&=&H_y e^{(d/2-|x|)/\lambda_y},\,\, H_z(x)=H_z e^{(d/2-|x|)/\lambda_z},\\
\lambda_y&=&\frac{c}{4e\eta\sqrt{2\pi J_3}},\\
\nonumber
\lambda_{z,L(R)}&=&\frac{c}{4e\eta}\sqrt{\frac{J_1+J_4\cos{2\alpha_{L(R)}}}{\!2\pi\!\left[J_1^2\!-\!J_4^2(1\!-\!\sin^2{2\alpha_{L(R)}}\sin^2{2\delta_{L(R)}})\right]}},
\end{eqnarray}
where $\lambda_{y,z}$ are corresponding London penetration depths ($\lambda_{y,L}=\lambda_{y,R}=\lambda_{y}$). In the considered case the components of the field decay independently. Similar to the $z$-junction, we can estimate the effective thickness of the $x$-junction as $d_i= d+\lambda_{i,L}+\lambda_{i,R}$ ($i=y,z$), $d_y$ corresponds to the applied field along $y$ axis and $d_z$ to the field along $z$ axis. 

In the case of the $x$-junction, the Josephson current has the component along the contact $j_y(x)$, Eq.~\eqref{jy_x}. According Eqs.~\eqref{London_current}, this current induces a $z$-component of the magnetic field $H_z \propto j_y'(x)$. In general, this field is not negligibly small since $j_y'(x) \propto j_y/d$. To avoid this difficulty we consider here only the purely nematic superconductors ($\delta_{R.L}=0$) and the short junction in a sufficiently large magnetic field, when the magnetic field in the junction is equal to the applied magnetic field. Similar to the case of the $z$-junction we can write the conditions of applicability of such an approach as [see Eqs.~\eqref{j_x_theta} and \eqref{lambda_J_z}] $H_i\gg \Phi_0/2\pi\lambda_{J,i}d_i$ and $\lambda_i\gg \lambda_{J,i}$, where
\begin{equation*}
\lambda_{J,i}=\sqrt{\frac{c\Phi_0}{8\pi^2d_ij_{c1}F(\alpha_R,\alpha_L)}}, 
\end{equation*}
\begin{equation}\label{f(RL)}
F(\alpha_R,\alpha_L)=\cos {\alpha_L} \cos {\alpha_R}+\frac{j_{c2}}{j_{c1}}\sin {\alpha_L} \sin {\alpha_R}.
\end{equation}

Now we can derive the dependence of the maximum Josephson current on the magnetic field in the case of the $x$-junction. We present here the results for two different orientations of the applied magnetic field, along $y$ and $z$ axes. Following the same procedure as in the previous subsection, we readily obtain
\begin{equation}\label{phase_A_x}
\nabla_i\theta=\frac{2\pi}{\Phi_0}\int_{-\infty}^{+\infty}\nabla_x A_i dx, 
\end{equation}
where $i=y,z$, $H_y=-\nabla_xA_z$ and $H_z=\nabla_xA_y$. After integration over $x$, we derive
\begin{eqnarray}\label{H_theta_x}
\nonumber
\nabla_y\theta=\frac{2\pi d_y}{\Phi_0}H_z, \quad \mathbf{H}_0=(0,0,H_z),\\
\nabla_z\theta=-\frac{2\pi d_z}{\Phi_0}H_y, \quad \mathbf{H}_0=(0,H_y,0).
\end{eqnarray}
These equations relates the phase difference on the junction and the applied magnetic field. Similar to the case of the $z$-junction, we can derive the dependence of the maximum current on the captured flux in the form 
\begin{eqnarray}\label{x_Imax}
\frac{I_{x\,max}(\Phi_i)}{I_{c1x}}=F(\alpha_R,\alpha_L)\,\frac{\sin{(\pi\Phi_i/\Phi_0)}}{\pi\Phi_i/\Phi_0},
\end{eqnarray}
where $i=y,z$ and $I_{c1x}=j_{c1}L$.

\section{Discussion}\label{Discussion}

One of the key result that we present in this work is that the superconductors with the nematic superconductivity in $E_u$ representation have off-diagonal components $K_{xy}$ in the Meissner kernel, see Eq.~\eqref{London_current}. As a result, we observe the anomalous Josephson Hall effect in the $x$-junction, that is, the supercurrent flowing along the contact, see Eq.~\eqref{jy_x} and Fig.~\ref{Figjyx}. We believe that the non-diagonal Meissner kernel can give rise to other peculiarities in electromagnetic properties of the Josephson junctions.    

The current in the junction is controlled by the vector order parameter at each side of the contact $\vec{\eta}=(\eta_{1R(L)},\eta_{1R(L)})$. It is convenient to parametrize the order parameter as $\vec{\eta}=e^{i\varphi}\eta(\cos \alpha e^{i\delta/2},\sin \alpha e^{-i\delta/2})$, where nematicity direction $\alpha$ and phase difference between the order parameters components $\delta$ determine the vector properties of the order parameter. When $\delta=0$, the nematicity direction $\alpha$ indicates direction of the order parameter vector in the coordinate space $(x,y)$. The orientation of the order parameter vectors to the contact significantly affects the current in the junction and we can control the Josephson current by tuning the nematicity in the bulk of the superconductors. The simplest strategy to control nematicity direction is the rotation of the superconductor in $(x,y)$ plane. Experiments show that large samples of the doped topological insulators are typically multidomain with different orientations of the nematicity vector in each domain. Weak links between different domains are natural objects to study the influence of $\alpha_{R, L}$ on the Josephson current. The control of the nematicity angle $\alpha$ by the external strain was demonstrated experimentally, see, e.g., Refs.~\onlinecite{Kostylev2020,Kuntsevich2019}. Thus, we can hope that the variation of the nematicity direction by the applied strain is a feasible task, which opens new possibilities for superconducting devices. 

Another parameter that controls the current is the phase difference between order parameters components $\delta$. In the case of non-zero $\delta$ an additional cosine term in the current-phase relations arises, see Eqs.~ \eqref{j_z_theta} and \eqref{j_x_theta}. For the pure nematic superconductor, this term vanishes, $\delta=0$, while for the pure chiral superconductors $\delta=\pi/2$ and the cosine is of importance. Theoretical analysis shows that the chiral superconducting phase can exist in the systems with an open Fermi surface~\cite{Venderbos2016_2, Khokhlov2021a}, in thin films~\cite{Chirolli2018} or under magnetic doping of the nematic superconductor~\cite{Chirolli2017, Yuan2017}. It is argued that in some experiments such a chiral phase with a finite magnetization has been observed~\cite{Neha2019, Qiu2015}. However, it is debated in other work~\cite{Das2020}. As we show in our previous work, the non-zero $\delta$ can be caused in the nematic phase by the external magnetization~\cite{Akzyanov2020_2}. The value of $\delta$ induced by the magnetization is small~\cite{Khokhlov2021a}. However, measurements of the currents in the Josephson junctions can be performed with high accuracy. We believe that the observation of the cosine term in the current that is caused by the magnetization is experimentally possible.

We observe that in the $x$-junction the Josephson Hall current along the junction, $j_y$, can be generated as well as the current across the junction, $j_x$. In some cases $j_y$ can be larger than $j_x$. Consider a junction made of two pure nematic superconductors with the order parameters $\vec{\eta_L}=\eta(1,0)$ and $\vec{\eta_R}=\eta e^{i\theta}(0,1)$ that orthogonal to each other and one of them is parallel to the contact plane. In this case longitudinal current vanishes $j_x=0$ [see Eq.~\eqref{j_x_theta}] but transverse current is non-zero $j_y=j_{cy} \sin \theta$ [see Eq.~\eqref{Hall_current}]. 

An analysis of Eq.~\eqref{jy_x} shows that under definite conditions the Josephson Hall current does not necessary decay exponentially with the increase of the junction thickness $d$ if $\delta_{L}$ and/or $\delta_{R}$ is non-zero. In the leading approximation with respect to $\kappa_i d \gg 1$, we obtain that $d\bar{j}_y \propto J_4\,\textrm{Im} (\eta_{1L}^* \eta_{2L}-\eta_{1R}^* \eta_{2R})$, which is non-zero for $d \rightarrow +\infty$. For example, such a situation takes place when $\delta\neq 0$ only in one side of the junction, which can be due to its magnetization or when one side of the junction is the chiral superconductor $\vec{\eta_L}\propto(1, i)$ and another side is nematic. It is worth mentioning that the Hall current flows even we consider a single boundary of the chiral superconductor, e.g. $\vec{\eta_R} =(0,0)$. Moreover, if $\delta$ in the superconductor is non-zero, then a finite current flows along the edge, $d\bar{j}_y\propto J_4 \eta_L^2 \sin 2\alpha_L \sin \delta_L$. Note, that the similar type of superconducting anomalous Hall effect has been predicted for a chiral p-wave superconductor~\cite{Furusaki2001}.

In Refs.~\cite{Wang2011, Vedyayev2013, Yokoyama2015, MatosAbiague2015, Dang2015, Mironov2017, Malshukov2019, Costa2020, Maistrenko2021} the anomalous Josephson Hall effect have been considered in the materials with a finite magnetization. This magnetization brings asymmetry to the Andreev bound states spectra that results in a finite Josephson Hall current. In our case, this supercurrent exist without magnetization. Thus, we establish that the anomalous Josephson Hall effect can be realized without any magnetization while time-reversal symmetry breaking occurs due to the phase difference across the contact.

One of the distinct features of London equations is the lack of gauge invariance for the current that leads to the lack of the charge conservation~\cite{Nambu1960,schriefferbook}. Usually, it is not a problem when the Meissner kernel is diagonal since we can always choose the proper gauge for a vector potential where the charge is locally conserved. In our case, we get formally that the charge is not conserved in case of the the $x$-junction if $\theta=\theta(y)$, which is the case for $z$ direction of the magnetic field. This drawback cannot be cured in a GL formalism. Instead, we should perform microscopic calculations of the current that include vertex corrections for a current operator that arises due to mean-field self-energy and Coulomb interaction. After that a global gauge invariance is restored and reliable results are obtained for any gauge of the vector potenial~\cite{Nambu1960,schriefferbook}. In the case of the anomalous superconducting Hall effect restoring the gauge invariance leads to the large additional quasiparticle contribution to the edge current~\cite{Stone2004}. It means that the anomalous Hall supercurrent is significantly renormalized by the vertex corrections. Thus, we expect a strong renormalization by the vertex corrections of the Josephson Hall current $j_y$ in the $x$-contact as well. However, microscopic calculations that include vertex corrections are beyond the scope of this paper. 

In conclusion, we study electromagnetic properties of Josephson junction betwee two nematic superconductors in the Ginzburg-Landau approach. We derive the London equations for the nematic superconductor with odd $E_u$ pairing using the second GL equation. We observe that the Meissner kernel has off-diagonal components. Using this result, we obtain current-phase relations for different orientations of the junction plane, crystallographic axes of the sample, and the nematicity vector. We show that the anomalous Josephson Hall effect can be observed in such in the absenct of magnetization. We calculate the magnetic field dependence of the maximum current through the junction. We show that the period of the maximum current oscillations with the captured in the junction magnetic flux depends on the geometry of the junction, direction of the magnetic field, and the nematicity vector.

\bibliography{vortex}

\begin{thebibliography}{63}%
\makeatletter
\providecommand \@ifxundefined [1]{%
 \@ifx{#1\undefined}
}%
\providecommand \@ifnum [1]{%
 \ifnum #1\expandafter \@firstoftwo
 \else \expandafter \@secondoftwo
 \fi
}%
\providecommand \@ifx [1]{%
 \ifx #1\expandafter \@firstoftwo
 \else \expandafter \@secondoftwo
 \fi
}%
\providecommand \natexlab [1]{#1}%
\providecommand \enquote  [1]{``#1''}%
\providecommand \bibnamefont  [1]{#1}%
\providecommand \bibfnamefont [1]{#1}%
\providecommand \citenamefont [1]{#1}%
\providecommand \href@noop [0]{\@secondoftwo}%
\providecommand \href [0]{\begingroup \@sanitize@url \@href}%
\providecommand \@href[1]{\@@startlink{#1}\@@href}%
\providecommand \@@href[1]{\endgroup#1\@@endlink}%
\providecommand \@sanitize@url [0]{\catcode `\\12\catcode `\$12\catcode
  `\&12\catcode `\#12\catcode `\^12\catcode `\_12\catcode `\%12\relax}%
\providecommand \@@startlink[1]{}%
\providecommand \@@endlink[0]{}%
\providecommand \url  [0]{\begingroup\@sanitize@url \@url }%
\providecommand \@url [1]{\endgroup\@href {#1}{\urlprefix }}%
\providecommand \urlprefix  [0]{URL }%
\providecommand \Eprint [0]{\href }%
\providecommand \doibase [0]{https://doi.org/}%
\providecommand \selectlanguage [0]{\@gobble}%
\providecommand \bibinfo  [0]{\@secondoftwo}%
\providecommand \bibfield  [0]{\@secondoftwo}%
\providecommand \translation [1]{[#1]}%
\providecommand \BibitemOpen [0]{}%
\providecommand \bibitemStop [0]{}%
\providecommand \bibitemNoStop [0]{.\EOS\space}%
\providecommand \EOS [0]{\spacefactor3000\relax}%
\providecommand \BibitemShut  [1]{\csname bibitem#1\endcsname}%
\let\auto@bib@innerbib\@empty
%</preamble>
\bibitem [{\citenamefont {Barone}\ and\ \citenamefont
  {Paterno}(1982)}]{barone1982}%
  \BibitemOpen
  \bibfield  {author} {\bibinfo {author} {\bibfnamefont {A.}~\bibnamefont
  {Barone}}\ and\ \bibinfo {author} {\bibfnamefont {G.}~\bibnamefont
  {Paterno}},\ }\href@noop {} {\emph {\bibinfo {title} {Physics and
  applications of the Josephson effect}}},\ Vol.~\bibinfo {volume} {1}\
  (\bibinfo  {publisher} {Wiley Online Library},\ \bibinfo {year}
  {1982})\BibitemShut {NoStop}%
\bibitem [{\citenamefont {Tinkham}(2004)}]{Tinkham2004}%
  \BibitemOpen
  \bibfield  {author} {\bibinfo {author} {\bibfnamefont {M.}~\bibnamefont
  {Tinkham}},\ }\href {http://www.worldcat.org/isbn/0486435032} {\emph
  {\bibinfo {title} {Introduction to Superconductivity}}},\ \bibinfo {edition}
  {2nd}\ ed.\ (\bibinfo  {publisher} {Dover Publications},\ \bibinfo {year}
  {2004})\BibitemShut {NoStop}%
\bibitem [{\citenamefont {Likharev}(1979)}]{Likharev1979}%
  \BibitemOpen
  \bibfield  {author} {\bibinfo {author} {\bibfnamefont {K.~K.}\ \bibnamefont
  {Likharev}},\ }\bibfield  {title} {\bibinfo {title} {Superconducting weak
  links},\ }\href {https://doi.org/10.1103/RevModPhys.51.101} {\bibfield
  {journal} {\bibinfo  {journal} {Rev. Mod. Phys.}\ }\textbf {\bibinfo {volume}
  {51}},\ \bibinfo {pages} {101} (\bibinfo {year} {1979})}\BibitemShut
  {NoStop}%
\bibitem [{\citenamefont {Tafuri}(2019)}]{tafuri2019}%
  \BibitemOpen
  \bibfield  {author} {\bibinfo {author} {\bibfnamefont {F.}~\bibnamefont
  {Tafuri}},\ }\href@noop {} {\emph {\bibinfo {title} {Fundamentals and
  frontiers of the Josephson effect}}},\ Vol.\ \bibinfo {volume} {286}\
  (\bibinfo  {publisher} {Springer Nature},\ \bibinfo {year}
  {2019})\BibitemShut {NoStop}%
\bibitem [{\citenamefont {Mel'nikov}\ \emph {et~al.}(2021)\citenamefont
  {Mel'nikov}, \citenamefont {Mironov}, \citenamefont {Samokhvalov},\ and\
  \citenamefont {Buzdin}}]{Melnikov2021}%
  \BibitemOpen
  \bibfield  {author} {\bibinfo {author} {\bibfnamefont {A.~S.}\ \bibnamefont
  {Mel'nikov}}, \bibinfo {author} {\bibfnamefont {S.~V.}\ \bibnamefont
  {Mironov}}, \bibinfo {author} {\bibfnamefont {A.~V.}\ \bibnamefont
  {Samokhvalov}},\ and\ \bibinfo {author} {\bibfnamefont {A.~I.}\ \bibnamefont
  {Buzdin}},\ }\bibfield  {title} {\bibinfo {title} {Superconducting
  spintronics: state of the art and perspectives},\ }\href@noop {} {\bibfield
  {journal} {\bibinfo  {journal} {Physics--Uspekhi}\ ,\ \bibinfo {pages} {64}}
  (\bibinfo {year} {2021})}\BibitemShut {NoStop}%
\bibitem [{\citenamefont {Wang}\ \emph {et~al.}(2011)\citenamefont {Wang},
  \citenamefont {Hao}, \citenamefont {Yang},\ and\ \citenamefont
  {Chan}}]{Wang2011}%
  \BibitemOpen
  \bibfield  {author} {\bibinfo {author} {\bibfnamefont {J.}~\bibnamefont
  {Wang}}, \bibinfo {author} {\bibfnamefont {L.}~\bibnamefont {Hao}}, \bibinfo
  {author} {\bibfnamefont {Y.~H.}\ \bibnamefont {Yang}},\ and\ \bibinfo
  {author} {\bibfnamefont {K.~S.}\ \bibnamefont {Chan}},\ }\bibfield  {title}
  {\bibinfo {title} {Josephson hall current in a noncentrosymmetric
  superconductor/ferromagnet/superconductor junction},\ }\href
  {https://doi.org/10.1063/1.3666055} {\bibfield  {journal} {\bibinfo
  {journal} {Journal of Applied Physics}\ }\textbf {\bibinfo {volume} {110}},\
  \bibinfo {pages} {113717} (\bibinfo {year} {2011})},\ \Eprint
  {https://arxiv.org/abs/https://doi.org/10.1063/1.3666055}
  {https://doi.org/10.1063/1.3666055} \BibitemShut {NoStop}%
\bibitem [{\citenamefont {Vedyayev}\ \emph {et~al.}(2013)\citenamefont
  {Vedyayev}, \citenamefont {Ryzhanova}, \citenamefont {Strelkov},\ and\
  \citenamefont {Dieny}}]{Vedyayev2013}%
  \BibitemOpen
  \bibfield  {author} {\bibinfo {author} {\bibfnamefont {A.}~\bibnamefont
  {Vedyayev}}, \bibinfo {author} {\bibfnamefont {N.}~\bibnamefont {Ryzhanova}},
  \bibinfo {author} {\bibfnamefont {N.}~\bibnamefont {Strelkov}},\ and\
  \bibinfo {author} {\bibfnamefont {B.}~\bibnamefont {Dieny}},\ }\bibfield
  {title} {\bibinfo {title} {Spontaneous anomalous and spin hall effects due to
  spin-orbit scattering of evanescent wave functions in magnetic tunnel
  junctions},\ }\href {https://doi.org/10.1103/PhysRevLett.110.247204}
  {\bibfield  {journal} {\bibinfo  {journal} {Phys. Rev. Lett.}\ }\textbf
  {\bibinfo {volume} {110}},\ \bibinfo {pages} {247204} (\bibinfo {year}
  {2013})}\BibitemShut {NoStop}%
\bibitem [{\citenamefont {Yokoyama}(2015)}]{Yokoyama2015}%
  \BibitemOpen
  \bibfield  {author} {\bibinfo {author} {\bibfnamefont {T.}~\bibnamefont
  {Yokoyama}},\ }\bibfield  {title} {\bibinfo {title} {Anomalous josephson hall
  effect in magnet/triplet superconductor junctions},\ }\href
  {https://doi.org/10.1103/PhysRevB.92.174513} {\bibfield  {journal} {\bibinfo
  {journal} {Phys. Rev. B}\ }\textbf {\bibinfo {volume} {92}},\ \bibinfo
  {pages} {174513} (\bibinfo {year} {2015})}\BibitemShut {NoStop}%
\bibitem [{\citenamefont {Matos-Abiague}\ and\ \citenamefont
  {Fabian}(2015)}]{MatosAbiague2015}%
  \BibitemOpen
  \bibfield  {author} {\bibinfo {author} {\bibfnamefont {A.}~\bibnamefont
  {Matos-Abiague}}\ and\ \bibinfo {author} {\bibfnamefont {J.}~\bibnamefont
  {Fabian}},\ }\bibfield  {title} {\bibinfo {title} {Tunneling anomalous and
  spin hall effects},\ }\href {https://doi.org/10.1103/PhysRevLett.115.056602}
  {\bibfield  {journal} {\bibinfo  {journal} {Phys. Rev. Lett.}\ }\textbf
  {\bibinfo {volume} {115}},\ \bibinfo {pages} {056602} (\bibinfo {year}
  {2015})}\BibitemShut {NoStop}%
\bibitem [{\citenamefont {Dang}\ \emph {et~al.}(2015)\citenamefont {Dang},
  \citenamefont {Jaffr\`es}, \citenamefont {Hoai~Nguyen},\ and\ \citenamefont
  {Drouhin}}]{Dang2015}%
  \BibitemOpen
  \bibfield  {author} {\bibinfo {author} {\bibfnamefont {T.~H.}\ \bibnamefont
  {Dang}}, \bibinfo {author} {\bibfnamefont {H.}~\bibnamefont {Jaffr\`es}},
  \bibinfo {author} {\bibfnamefont {T.~L.}\ \bibnamefont {Hoai~Nguyen}},\ and\
  \bibinfo {author} {\bibfnamefont {H.-J.}\ \bibnamefont {Drouhin}},\
  }\bibfield  {title} {\bibinfo {title} {Giant forward-scattering asymmetry and
  anomalous tunnel hall effect at spin-orbit-split and exchange-split
  interfaces},\ }\href {https://doi.org/10.1103/PhysRevB.92.060403} {\bibfield
  {journal} {\bibinfo  {journal} {Phys. Rev. B}\ }\textbf {\bibinfo {volume}
  {92}},\ \bibinfo {pages} {060403} (\bibinfo {year} {2015})}\BibitemShut
  {NoStop}%
\bibitem [{\citenamefont {Mironov}\ and\ \citenamefont
  {Buzdin}(2017)}]{Mironov2017}%
  \BibitemOpen
  \bibfield  {author} {\bibinfo {author} {\bibfnamefont {S.}~\bibnamefont
  {Mironov}}\ and\ \bibinfo {author} {\bibfnamefont {A.}~\bibnamefont
  {Buzdin}},\ }\bibfield  {title} {\bibinfo {title} {Spontaneous currents in
  superconducting systems with strong spin-orbit coupling},\ }\href
  {https://doi.org/10.1103/PhysRevLett.118.077001} {\bibfield  {journal}
  {\bibinfo  {journal} {Phys. Rev. Lett.}\ }\textbf {\bibinfo {volume} {118}},\
  \bibinfo {pages} {077001} (\bibinfo {year} {2017})}\BibitemShut {NoStop}%
\bibitem [{\citenamefont {Mal'shukov}(2019)}]{Malshukov2019}%
  \BibitemOpen
  \bibfield  {author} {\bibinfo {author} {\bibfnamefont {A.~G.}\ \bibnamefont
  {Mal'shukov}},\ }\bibfield  {title} {\bibinfo {title} {Ac anomalous hall
  effect in topological insulator josephson junctions},\ }\href
  {https://doi.org/10.1103/PhysRevB.100.035301} {\bibfield  {journal} {\bibinfo
   {journal} {Phys. Rev. B}\ }\textbf {\bibinfo {volume} {100}},\ \bibinfo
  {pages} {035301} (\bibinfo {year} {2019})}\BibitemShut {NoStop}%
\bibitem [{\citenamefont {Costa}\ and\ \citenamefont
  {Fabian}(2020)}]{Costa2020}%
  \BibitemOpen
  \bibfield  {author} {\bibinfo {author} {\bibfnamefont {A.}~\bibnamefont
  {Costa}}\ and\ \bibinfo {author} {\bibfnamefont {J.}~\bibnamefont {Fabian}},\
  }\bibfield  {title} {\bibinfo {title} {Anomalous josephson hall effect charge
  and transverse spin currents in
  superconductor/ferromagnetic-insulator/superconductor junctions},\ }\href
  {https://doi.org/10.1103/PhysRevB.101.104508} {\bibfield  {journal} {\bibinfo
   {journal} {Phys. Rev. B}\ }\textbf {\bibinfo {volume} {101}},\ \bibinfo
  {pages} {104508} (\bibinfo {year} {2020})}\BibitemShut {NoStop}%
\bibitem [{\citenamefont {Maistrenko}\ \emph {et~al.}(2021)\citenamefont
  {Maistrenko}, \citenamefont {Scharf}, \citenamefont {Manske},\ and\
  \citenamefont {Hankiewicz}}]{Maistrenko2021}%
  \BibitemOpen
  \bibfield  {author} {\bibinfo {author} {\bibfnamefont {O.}~\bibnamefont
  {Maistrenko}}, \bibinfo {author} {\bibfnamefont {B.}~\bibnamefont {Scharf}},
  \bibinfo {author} {\bibfnamefont {D.}~\bibnamefont {Manske}},\ and\ \bibinfo
  {author} {\bibfnamefont {E.~M.}\ \bibnamefont {Hankiewicz}},\ }\bibfield
  {title} {\bibinfo {title} {Planar josephson hall effect in topological
  josephson junctions},\ }\href {https://doi.org/10.1103/PhysRevB.103.054508}
  {\bibfield  {journal} {\bibinfo  {journal} {Phys. Rev. B}\ }\textbf {\bibinfo
  {volume} {103}},\ \bibinfo {pages} {054508} (\bibinfo {year}
  {2021})}\BibitemShut {NoStop}%
\bibitem [{\citenamefont {Nagaosa}\ \emph {et~al.}(2010)\citenamefont
  {Nagaosa}, \citenamefont {Sinova}, \citenamefont {Onoda}, \citenamefont
  {MacDonald},\ and\ \citenamefont {Ong}}]{Nagaosa2010}%
  \BibitemOpen
  \bibfield  {author} {\bibinfo {author} {\bibfnamefont {N.}~\bibnamefont
  {Nagaosa}}, \bibinfo {author} {\bibfnamefont {J.}~\bibnamefont {Sinova}},
  \bibinfo {author} {\bibfnamefont {S.}~\bibnamefont {Onoda}}, \bibinfo
  {author} {\bibfnamefont {A.~H.}\ \bibnamefont {MacDonald}},\ and\ \bibinfo
  {author} {\bibfnamefont {N.~P.}\ \bibnamefont {Ong}},\ }\bibfield  {title}
  {\bibinfo {title} {Anomalous hall effect},\ }\href
  {https://doi.org/10.1103/RevModPhys.82.1539} {\bibfield  {journal} {\bibinfo
  {journal} {Rev. Mod. Phys.}\ }\textbf {\bibinfo {volume} {82}},\ \bibinfo
  {pages} {1539} (\bibinfo {year} {2010})}\BibitemShut {NoStop}%
\bibitem [{\citenamefont {Hor}\ \emph {et~al.}(2010)\citenamefont {Hor},
  \citenamefont {Williams}, \citenamefont {Checkelsky}, \citenamefont
  {Roushan}, \citenamefont {Seo}, \citenamefont {Xu}, \citenamefont
  {Zandbergen}, \citenamefont {Yazdani}, \citenamefont {Ong},\ and\
  \citenamefont {Cava}}]{Hor2010}%
  \BibitemOpen
  \bibfield  {author} {\bibinfo {author} {\bibfnamefont {Y.~S.}\ \bibnamefont
  {Hor}}, \bibinfo {author} {\bibfnamefont {A.~J.}\ \bibnamefont {Williams}},
  \bibinfo {author} {\bibfnamefont {J.~G.}\ \bibnamefont {Checkelsky}},
  \bibinfo {author} {\bibfnamefont {P.}~\bibnamefont {Roushan}}, \bibinfo
  {author} {\bibfnamefont {J.}~\bibnamefont {Seo}}, \bibinfo {author}
  {\bibfnamefont {Q.}~\bibnamefont {Xu}}, \bibinfo {author} {\bibfnamefont
  {H.~W.}\ \bibnamefont {Zandbergen}}, \bibinfo {author} {\bibfnamefont
  {A.}~\bibnamefont {Yazdani}}, \bibinfo {author} {\bibfnamefont {N.~P.}\
  \bibnamefont {Ong}},\ and\ \bibinfo {author} {\bibfnamefont {R.~J.}\
  \bibnamefont {Cava}},\ }\bibfield  {title} {\bibinfo {title}
  {Superconductivity in ${\mathrm{cu}}_{x}{\mathrm{bi}}_{2}{\mathrm{se}}_{3}$
  and its implications for pairing in the undoped topological insulator},\
  }\href {https://link.aps.org/doi/10.1103/PhysRevLett.104.057001} {\bibfield
  {journal} {\bibinfo  {journal} {Phys. Rev. Lett.}\ }\textbf {\bibinfo
  {volume} {104}},\ \bibinfo {pages} {057001} (\bibinfo {year}
  {2010})}\BibitemShut {NoStop}%
\bibitem [{\citenamefont {Sasaki}\ \emph {et~al.}(2011)\citenamefont {Sasaki},
  \citenamefont {Kriener}, \citenamefont {Segawa}, \citenamefont {Yada},
  \citenamefont {Tanaka}, \citenamefont {Sato},\ and\ \citenamefont
  {Ando}}]{Sasaki2011}%
  \BibitemOpen
  \bibfield  {author} {\bibinfo {author} {\bibfnamefont {S.}~\bibnamefont
  {Sasaki}}, \bibinfo {author} {\bibfnamefont {M.}~\bibnamefont {Kriener}},
  \bibinfo {author} {\bibfnamefont {K.}~\bibnamefont {Segawa}}, \bibinfo
  {author} {\bibfnamefont {K.}~\bibnamefont {Yada}}, \bibinfo {author}
  {\bibfnamefont {Y.}~\bibnamefont {Tanaka}}, \bibinfo {author} {\bibfnamefont
  {M.}~\bibnamefont {Sato}},\ and\ \bibinfo {author} {\bibfnamefont
  {Y.}~\bibnamefont {Ando}},\ }\bibfield  {title} {\bibinfo {title}
  {Topological superconductivity in
  ${\mathrm{cu}}_{x}{\mathrm{bi}}_{2}{\mathrm{se}}_{3}$},\ }\href
  {https://doi.org/10.1103/PhysRevLett.107.217001} {\bibfield  {journal}
  {\bibinfo  {journal} {Phys. Rev. Lett.}\ }\textbf {\bibinfo {volume} {107}},\
  \bibinfo {pages} {217001} (\bibinfo {year} {2011})}\BibitemShut {NoStop}%
\bibitem [{\citenamefont {Kirzhner}\ \emph {et~al.}(2012)\citenamefont
  {Kirzhner}, \citenamefont {Lahoud}, \citenamefont {Chaska}, \citenamefont
  {Salman},\ and\ \citenamefont {Kanigel}}]{Kirzhner2012}%
  \BibitemOpen
  \bibfield  {author} {\bibinfo {author} {\bibfnamefont {T.}~\bibnamefont
  {Kirzhner}}, \bibinfo {author} {\bibfnamefont {E.}~\bibnamefont {Lahoud}},
  \bibinfo {author} {\bibfnamefont {K.~B.}\ \bibnamefont {Chaska}}, \bibinfo
  {author} {\bibfnamefont {Z.}~\bibnamefont {Salman}},\ and\ \bibinfo {author}
  {\bibfnamefont {A.}~\bibnamefont {Kanigel}},\ }\bibfield  {title} {\bibinfo
  {title} {Point-contact spectroscopy of cu${}_{0.2}$bi${}_{2}$se${}_{3}$
  single crystals},\ }\href
  {https://link.aps.org/doi/10.1103/PhysRevB.86.064517} {\bibfield  {journal}
  {\bibinfo  {journal} {Phys. Rev. B}\ }\textbf {\bibinfo {volume} {86}},\
  \bibinfo {pages} {064517} (\bibinfo {year} {2012})}\BibitemShut {NoStop}%
\bibitem [{\citenamefont {Kawai}\ \emph {et~al.}(2020)\citenamefont {Kawai},
  \citenamefont {Wang}, \citenamefont {Kandori}, \citenamefont {Honoki},
  \citenamefont {Matano}, \citenamefont {Kambe},\ and\ \citenamefont {qing
  Zheng}}]{Kawai2020}%
  \BibitemOpen
  \bibfield  {author} {\bibinfo {author} {\bibfnamefont {T.}~\bibnamefont
  {Kawai}}, \bibinfo {author} {\bibfnamefont {C.~G.}\ \bibnamefont {Wang}},
  \bibinfo {author} {\bibfnamefont {Y.}~\bibnamefont {Kandori}}, \bibinfo
  {author} {\bibfnamefont {Y.}~\bibnamefont {Honoki}}, \bibinfo {author}
  {\bibfnamefont {K.}~\bibnamefont {Matano}}, \bibinfo {author} {\bibfnamefont
  {T.}~\bibnamefont {Kambe}},\ and\ \bibinfo {author} {\bibfnamefont
  {G.}~\bibnamefont {qing Zheng}},\ }\bibfield  {title} {\bibinfo {title}
  {Direction and symmetry transition of the vector order parameter in
  topological superconductors {CuxBi}2se3},\ }\bibfield  {journal} {\bibinfo
  {journal} {Nature Communications}\ }\textbf {\bibinfo {volume} {11}},\ \href
  {https://doi.org/10.1038/s41467-019-14126-w} {10.1038/s41467-019-14126-w}
  (\bibinfo {year} {2020})\BibitemShut {NoStop}%
\bibitem [{\citenamefont {Yonezawa}\ \emph {et~al.}(2016)\citenamefont
  {Yonezawa}, \citenamefont {Tajiri}, \citenamefont {Nakata}, \citenamefont
  {Nagai}, \citenamefont {Wang}, \citenamefont {Segawa}, \citenamefont {Ando},\
  and\ \citenamefont {Maeno}}]{Yonezawa2016}%
  \BibitemOpen
  \bibfield  {author} {\bibinfo {author} {\bibfnamefont {S.}~\bibnamefont
  {Yonezawa}}, \bibinfo {author} {\bibfnamefont {K.}~\bibnamefont {Tajiri}},
  \bibinfo {author} {\bibfnamefont {S.}~\bibnamefont {Nakata}}, \bibinfo
  {author} {\bibfnamefont {Y.}~\bibnamefont {Nagai}}, \bibinfo {author}
  {\bibfnamefont {Z.}~\bibnamefont {Wang}}, \bibinfo {author} {\bibfnamefont
  {K.}~\bibnamefont {Segawa}}, \bibinfo {author} {\bibfnamefont
  {Y.}~\bibnamefont {Ando}},\ and\ \bibinfo {author} {\bibfnamefont
  {Y.}~\bibnamefont {Maeno}},\ }\bibfield  {title} {\bibinfo {title}
  {Thermodynamic evidence for nematic superconductivity in {CuxBi}2se3},\
  }\href {https://doi.org/10.1038/nphys3907} {\bibfield  {journal} {\bibinfo
  {journal} {Nature Physics}\ }\textbf {\bibinfo {volume} {13}},\ \bibinfo
  {pages} {123} (\bibinfo {year} {2016})}\BibitemShut {NoStop}%
\bibitem [{\citenamefont {Tao}\ \emph {et~al.}(2018)\citenamefont {Tao},
  \citenamefont {Yan}, \citenamefont {Liu}, \citenamefont {Wang}, \citenamefont
  {Ando}, \citenamefont {Wang}, \citenamefont {Zhang},\ and\ \citenamefont
  {Feng}}]{Tao2018}%
  \BibitemOpen
  \bibfield  {author} {\bibinfo {author} {\bibfnamefont {R.}~\bibnamefont
  {Tao}}, \bibinfo {author} {\bibfnamefont {Y.-J.}\ \bibnamefont {Yan}},
  \bibinfo {author} {\bibfnamefont {X.}~\bibnamefont {Liu}}, \bibinfo {author}
  {\bibfnamefont {Z.-W.}\ \bibnamefont {Wang}}, \bibinfo {author}
  {\bibfnamefont {Y.}~\bibnamefont {Ando}}, \bibinfo {author} {\bibfnamefont
  {Q.-H.}\ \bibnamefont {Wang}}, \bibinfo {author} {\bibfnamefont
  {T.}~\bibnamefont {Zhang}},\ and\ \bibinfo {author} {\bibfnamefont {D.-L.}\
  \bibnamefont {Feng}},\ }\bibfield  {title} {\bibinfo {title} {Direct
  visualization of the nematic superconductivity in
  ${\mathrm{cu}}_{x}{\mathrm{bi}}_{2}{\mathrm{se}}_{3}$},\ }\href
  {https://doi.org/10.1103/PhysRevX.8.041024} {\bibfield  {journal} {\bibinfo
  {journal} {Phys. Rev. X}\ }\textbf {\bibinfo {volume} {8}},\ \bibinfo {pages}
  {041024} (\bibinfo {year} {2018})}\BibitemShut {NoStop}%
\bibitem [{\citenamefont {Matano}\ \emph {et~al.}(2016)\citenamefont {Matano},
  \citenamefont {Kriener}, \citenamefont {Segawa}, \citenamefont {Ando},\ and\
  \citenamefont {qing Zheng}}]{Matano2016}%
  \BibitemOpen
  \bibfield  {author} {\bibinfo {author} {\bibfnamefont {K.}~\bibnamefont
  {Matano}}, \bibinfo {author} {\bibfnamefont {M.}~\bibnamefont {Kriener}},
  \bibinfo {author} {\bibfnamefont {K.}~\bibnamefont {Segawa}}, \bibinfo
  {author} {\bibfnamefont {Y.}~\bibnamefont {Ando}},\ and\ \bibinfo {author}
  {\bibfnamefont {G.}~\bibnamefont {qing Zheng}},\ }\bibfield  {title}
  {\bibinfo {title} {Spin-rotation symmetry breaking in the superconducting
  state of {CuxBi}2se3},\ }\href {https://doi.org/10.1038/nphys3781} {\bibfield
   {journal} {\bibinfo  {journal} {Nature Physics}\ }\textbf {\bibinfo {volume}
  {12}},\ \bibinfo {pages} {852} (\bibinfo {year} {2016})}\BibitemShut
  {NoStop}%
\bibitem [{\citenamefont {Shruti}\ \emph {et~al.}(2015)\citenamefont {Shruti},
  \citenamefont {Maurya}, \citenamefont {Neha}, \citenamefont {Srivastava},\
  and\ \citenamefont {Patnaik}}]{Shruti2015}%
  \BibitemOpen
  \bibfield  {author} {\bibinfo {author} {\bibnamefont {Shruti}}, \bibinfo
  {author} {\bibfnamefont {V.~K.}\ \bibnamefont {Maurya}}, \bibinfo {author}
  {\bibfnamefont {P.}~\bibnamefont {Neha}}, \bibinfo {author} {\bibfnamefont
  {P.}~\bibnamefont {Srivastava}},\ and\ \bibinfo {author} {\bibfnamefont
  {S.}~\bibnamefont {Patnaik}},\ }\bibfield  {title} {\bibinfo {title}
  {Superconductivity by sr intercalation in the layered topological insulator
  ${\mathrm{bi}}_{2}{\mathrm{se}}_{3}$},\ }\href
  {https://doi.org/10.1103/PhysRevB.92.020506} {\bibfield  {journal} {\bibinfo
  {journal} {Phys. Rev. B}\ }\textbf {\bibinfo {volume} {92}},\ \bibinfo
  {pages} {020506(R)} (\bibinfo {year} {2015})}\BibitemShut {NoStop}%
\bibitem [{\citenamefont {Liu}\ \emph {et~al.}(2015)\citenamefont {Liu},
  \citenamefont {Yao}, \citenamefont {Shao}, \citenamefont {Zuo}, \citenamefont
  {Pi}, \citenamefont {Tan}, \citenamefont {Zhang},\ and\ \citenamefont
  {Zhang}}]{Liu2015}%
  \BibitemOpen
  \bibfield  {author} {\bibinfo {author} {\bibfnamefont {Z.}~\bibnamefont
  {Liu}}, \bibinfo {author} {\bibfnamefont {X.}~\bibnamefont {Yao}}, \bibinfo
  {author} {\bibfnamefont {J.}~\bibnamefont {Shao}}, \bibinfo {author}
  {\bibfnamefont {M.}~\bibnamefont {Zuo}}, \bibinfo {author} {\bibfnamefont
  {L.}~\bibnamefont {Pi}}, \bibinfo {author} {\bibfnamefont {S.}~\bibnamefont
  {Tan}}, \bibinfo {author} {\bibfnamefont {C.}~\bibnamefont {Zhang}},\ and\
  \bibinfo {author} {\bibfnamefont {Y.}~\bibnamefont {Zhang}},\ }\bibfield
  {title} {\bibinfo {title} {Superconductivity with topological surface state
  in {SrxBi}2se3},\ }\href {https://doi.org/10.1021/jacs.5b06815} {\bibfield
  {journal} {\bibinfo  {journal} {Journal of the American Chemical Society}\
  }\textbf {\bibinfo {volume} {137}},\ \bibinfo {pages} {10512} (\bibinfo
  {year} {2015})}\BibitemShut {NoStop}%
\bibitem [{\citenamefont {Kuntsevich}\ \emph {et~al.}(2018)\citenamefont
  {Kuntsevich}, \citenamefont {Bryzgalov}, \citenamefont {Prudkoglyad},
  \citenamefont {Martovitskii}, \citenamefont {Selivanov},\ and\ \citenamefont
  {Chizhevskii}}]{Kuntsevich2018}%
  \BibitemOpen
  \bibfield  {author} {\bibinfo {author} {\bibfnamefont {A.~Y.}\ \bibnamefont
  {Kuntsevich}}, \bibinfo {author} {\bibfnamefont {M.~A.}\ \bibnamefont
  {Bryzgalov}}, \bibinfo {author} {\bibfnamefont {V.~A.}\ \bibnamefont
  {Prudkoglyad}}, \bibinfo {author} {\bibfnamefont {V.~P.}\ \bibnamefont
  {Martovitskii}}, \bibinfo {author} {\bibfnamefont {Y.~G.}\ \bibnamefont
  {Selivanov}},\ and\ \bibinfo {author} {\bibfnamefont {E.~G.}\ \bibnamefont
  {Chizhevskii}},\ }\bibfield  {title} {\bibinfo {title} {Structural distortion
  behind the nematic superconductivity in sr x bi2se3},\ }\href
  {https://doi.org/10.1088/1367-2630/aae595} {\bibfield  {journal} {\bibinfo
  {journal} {New Journal of Physics}\ }\textbf {\bibinfo {volume} {20}},\
  \bibinfo {pages} {103022} (\bibinfo {year} {2018})}\BibitemShut {NoStop}%
\bibitem [{\citenamefont {Kuntsevich}\ \emph {et~al.}(2019)\citenamefont
  {Kuntsevich}, \citenamefont {Bryzgalov}, \citenamefont {Akzyanov},
  \citenamefont {Martovitskii}, \citenamefont {Rakhmanov},\ and\ \citenamefont
  {Selivanov}}]{Kuntsevich2019}%
  \BibitemOpen
  \bibfield  {author} {\bibinfo {author} {\bibfnamefont {A.~Y.}\ \bibnamefont
  {Kuntsevich}}, \bibinfo {author} {\bibfnamefont {M.~A.}\ \bibnamefont
  {Bryzgalov}}, \bibinfo {author} {\bibfnamefont {R.~S.}\ \bibnamefont
  {Akzyanov}}, \bibinfo {author} {\bibfnamefont {V.~P.}\ \bibnamefont
  {Martovitskii}}, \bibinfo {author} {\bibfnamefont {A.~L.}\ \bibnamefont
  {Rakhmanov}},\ and\ \bibinfo {author} {\bibfnamefont {Y.~G.}\ \bibnamefont
  {Selivanov}},\ }\bibfield  {title} {\bibinfo {title} {Strain-driven
  nematicity of odd-parity superconductivity in
  ${\mathrm{sr}}_{x}{\mathrm{bi}}_{2}{\mathrm{se}}_{3}$},\ }\href
  {https://doi.org/10.1103/PhysRevB.100.224509} {\bibfield  {journal} {\bibinfo
   {journal} {Phys. Rev. B}\ }\textbf {\bibinfo {volume} {100}},\ \bibinfo
  {pages} {224509} (\bibinfo {year} {2019})}\BibitemShut {NoStop}%
\bibitem [{\citenamefont {Pan}\ \emph {et~al.}(2016)\citenamefont {Pan},
  \citenamefont {Nikitin}, \citenamefont {Araizi}, \citenamefont {Huang},
  \citenamefont {Matsushita}, \citenamefont {Naka},\ and\ \citenamefont
  {de~Visser}}]{Pan2016}%
  \BibitemOpen
  \bibfield  {author} {\bibinfo {author} {\bibfnamefont {Y.}~\bibnamefont
  {Pan}}, \bibinfo {author} {\bibfnamefont {A.~M.}\ \bibnamefont {Nikitin}},
  \bibinfo {author} {\bibfnamefont {G.~K.}\ \bibnamefont {Araizi}}, \bibinfo
  {author} {\bibfnamefont {Y.~K.}\ \bibnamefont {Huang}}, \bibinfo {author}
  {\bibfnamefont {Y.}~\bibnamefont {Matsushita}}, \bibinfo {author}
  {\bibfnamefont {T.}~\bibnamefont {Naka}},\ and\ \bibinfo {author}
  {\bibfnamefont {A.}~\bibnamefont {de~Visser}},\ }\bibfield  {title} {\bibinfo
  {title} {Rotational symmetry breaking in the topological superconductor
  srxbi2se3 probed by upper-critical field experiments},\ }\href
  {https://doi.org/10.1038/srep28632} {\bibfield  {journal} {\bibinfo
  {journal} {Scientific Reports}\ }\textbf {\bibinfo {volume} {6}},\ \bibinfo
  {pages} {28632} (\bibinfo {year} {2016})}\BibitemShut {NoStop}%
\bibitem [{\citenamefont {Neha}\ \emph {et~al.}(2019)\citenamefont {Neha},
  \citenamefont {Biswas}, \citenamefont {Das},\ and\ \citenamefont
  {Patnaik}}]{Neha2019}%
  \BibitemOpen
  \bibfield  {author} {\bibinfo {author} {\bibfnamefont {P.}~\bibnamefont
  {Neha}}, \bibinfo {author} {\bibfnamefont {P.~K.}\ \bibnamefont {Biswas}},
  \bibinfo {author} {\bibfnamefont {T.}~\bibnamefont {Das}},\ and\ \bibinfo
  {author} {\bibfnamefont {S.}~\bibnamefont {Patnaik}},\ }\bibfield  {title}
  {\bibinfo {title} {Time-reversal symmetry breaking in topological
  superconductor ${\mathrm{sr}}_{0.1}{\mathrm{bi}}_{2}{\mathrm{se}}_{3}$},\
  }\href {https://doi.org/10.1103/PhysRevMaterials.3.074201} {\bibfield
  {journal} {\bibinfo  {journal} {Phys. Rev. Materials}\ }\textbf {\bibinfo
  {volume} {3}},\ \bibinfo {pages} {074201} (\bibinfo {year}
  {2019})}\BibitemShut {NoStop}%
\bibitem [{\citenamefont {Bannikov}\ \emph {et~al.}(2021)\citenamefont
  {Bannikov}, \citenamefont {Akzyanov}, \citenamefont {Zhurbina}, \citenamefont
  {Khaldeev}, \citenamefont {Selivanov}, \citenamefont {Zavyalov},
  \citenamefont {Rakhmanov},\ and\ \citenamefont {Kuntsevich}}]{Bannikov2021}%
  \BibitemOpen
  \bibfield  {author} {\bibinfo {author} {\bibfnamefont {M.~I.}\ \bibnamefont
  {Bannikov}}, \bibinfo {author} {\bibfnamefont {R.~S.}\ \bibnamefont
  {Akzyanov}}, \bibinfo {author} {\bibfnamefont {N.~K.}\ \bibnamefont
  {Zhurbina}}, \bibinfo {author} {\bibfnamefont {S.~I.}\ \bibnamefont
  {Khaldeev}}, \bibinfo {author} {\bibfnamefont {Y.~G.}\ \bibnamefont
  {Selivanov}}, \bibinfo {author} {\bibfnamefont {V.~V.}\ \bibnamefont
  {Zavyalov}}, \bibinfo {author} {\bibfnamefont {A.~L.}\ \bibnamefont
  {Rakhmanov}},\ and\ \bibinfo {author} {\bibfnamefont {A.~Y.}\ \bibnamefont
  {Kuntsevich}},\ }\bibfield  {title} {\bibinfo {title} {Breaking of
  ginzburg-landau description in the temperature dependence of the anisotropy
  in a nematic superconductor},\ }\href
  {https://doi.org/10.1103/PhysRevB.104.L220502} {\bibfield  {journal}
  {\bibinfo  {journal} {Phys. Rev. B}\ }\textbf {\bibinfo {volume} {104}},\
  \bibinfo {pages} {L220502} (\bibinfo {year} {2021})}\BibitemShut {NoStop}%
\bibitem [{\citenamefont {Qiu}\ \emph {et~al.}(2015)\citenamefont {Qiu},
  \citenamefont {Sanders}, \citenamefont {Dai}, \citenamefont {Medvedeva},
  \citenamefont {Wu}, \citenamefont {Ghaemi}, \citenamefont {Vojta},\ and\
  \citenamefont {Hor}}]{Qiu2015}%
  \BibitemOpen
  \bibfield  {author} {\bibinfo {author} {\bibfnamefont {Y.}~\bibnamefont
  {Qiu}}, \bibinfo {author} {\bibfnamefont {K.~N.}\ \bibnamefont {Sanders}},
  \bibinfo {author} {\bibfnamefont {J.}~\bibnamefont {Dai}}, \bibinfo {author}
  {\bibfnamefont {J.~E.}\ \bibnamefont {Medvedeva}}, \bibinfo {author}
  {\bibfnamefont {W.}~\bibnamefont {Wu}}, \bibinfo {author} {\bibfnamefont
  {P.}~\bibnamefont {Ghaemi}}, \bibinfo {author} {\bibfnamefont
  {T.}~\bibnamefont {Vojta}},\ and\ \bibinfo {author} {\bibfnamefont {Y.~S.}\
  \bibnamefont {Hor}},\ }\href@noop {} {\bibinfo {title} {Time reversal
  symmetry breaking superconductivity in topological materials}} (\bibinfo
  {year} {2015}),\ \Eprint {https://arxiv.org/abs/1512.03519} {arXiv:1512.03519
  [cond-mat.supr-con]} \BibitemShut {NoStop}%
\bibitem [{\citenamefont {Kurter}\ \emph {et~al.}(2018)\citenamefont {Kurter},
  \citenamefont {Finck}, \citenamefont {Huemiller}, \citenamefont {Medvedeva},
  \citenamefont {Weis}, \citenamefont {Atkinson}, \citenamefont {Qiu},
  \citenamefont {Shen}, \citenamefont {Lee}, \citenamefont {Vojta},
  \citenamefont {Ghaemi}, \citenamefont {Hor},\ and\ \citenamefont
  {Harlingen}}]{Kurter2018}%
  \BibitemOpen
  \bibfield  {author} {\bibinfo {author} {\bibfnamefont {C.}~\bibnamefont
  {Kurter}}, \bibinfo {author} {\bibfnamefont {A.~D.~K.}\ \bibnamefont
  {Finck}}, \bibinfo {author} {\bibfnamefont {E.~D.}\ \bibnamefont
  {Huemiller}}, \bibinfo {author} {\bibfnamefont {J.}~\bibnamefont
  {Medvedeva}}, \bibinfo {author} {\bibfnamefont {A.}~\bibnamefont {Weis}},
  \bibinfo {author} {\bibfnamefont {J.~M.}\ \bibnamefont {Atkinson}}, \bibinfo
  {author} {\bibfnamefont {Y.}~\bibnamefont {Qiu}}, \bibinfo {author}
  {\bibfnamefont {L.}~\bibnamefont {Shen}}, \bibinfo {author} {\bibfnamefont
  {S.~H.}\ \bibnamefont {Lee}}, \bibinfo {author} {\bibfnamefont
  {T.}~\bibnamefont {Vojta}}, \bibinfo {author} {\bibfnamefont
  {P.}~\bibnamefont {Ghaemi}}, \bibinfo {author} {\bibfnamefont {Y.~S.}\
  \bibnamefont {Hor}},\ and\ \bibinfo {author} {\bibfnamefont {D.~J.~V.}\
  \bibnamefont {Harlingen}},\ }\bibfield  {title} {\bibinfo {title}
  {Conductance spectroscopy of exfoliated thin flakes of {NbxBi}2se3},\ }\href
  {https://doi.org/10.1021/acs.nanolett.8b02954} {\bibfield  {journal}
  {\bibinfo  {journal} {Nano Letters}\ }\textbf {\bibinfo {volume} {19}},\
  \bibinfo {pages} {38} (\bibinfo {year} {2018})}\BibitemShut {NoStop}%
\bibitem [{\citenamefont {Asaba}\ \emph {et~al.}(2017)\citenamefont {Asaba},
  \citenamefont {Lawson}, \citenamefont {Tinsman}, \citenamefont {Chen},
  \citenamefont {Corbae}, \citenamefont {Li}, \citenamefont {Qiu},
  \citenamefont {Hor}, \citenamefont {Fu},\ and\ \citenamefont
  {Li}}]{Asaba2017}%
  \BibitemOpen
  \bibfield  {author} {\bibinfo {author} {\bibfnamefont {T.}~\bibnamefont
  {Asaba}}, \bibinfo {author} {\bibfnamefont {B.~J.}\ \bibnamefont {Lawson}},
  \bibinfo {author} {\bibfnamefont {C.}~\bibnamefont {Tinsman}}, \bibinfo
  {author} {\bibfnamefont {L.}~\bibnamefont {Chen}}, \bibinfo {author}
  {\bibfnamefont {P.}~\bibnamefont {Corbae}}, \bibinfo {author} {\bibfnamefont
  {G.}~\bibnamefont {Li}}, \bibinfo {author} {\bibfnamefont {Y.}~\bibnamefont
  {Qiu}}, \bibinfo {author} {\bibfnamefont {Y.~S.}\ \bibnamefont {Hor}},
  \bibinfo {author} {\bibfnamefont {L.}~\bibnamefont {Fu}},\ and\ \bibinfo
  {author} {\bibfnamefont {L.}~\bibnamefont {Li}},\ }\bibfield  {title}
  {\bibinfo {title} {Rotational symmetry breaking in a trigonal superconductor
  nb-doped ${\mathrm{bi}}_{2}{\mathrm{se}}_{3}$},\ }\href
  {https://doi.org/10.1103/PhysRevX.7.011009} {\bibfield  {journal} {\bibinfo
  {journal} {Phys. Rev. X}\ }\textbf {\bibinfo {volume} {7}},\ \bibinfo {pages}
  {011009} (\bibinfo {year} {2017})}\BibitemShut {NoStop}%
\bibitem [{\citenamefont {Das}\ \emph {et~al.}(2020)\citenamefont {Das},
  \citenamefont {Kobayashi}, \citenamefont {Smylie}, \citenamefont {Mielke},
  \citenamefont {Takahashi}, \citenamefont {Willa}, \citenamefont {Yin},
  \citenamefont {Welp}, \citenamefont {Hasan}, \citenamefont {Amato},
  \citenamefont {Luetkens},\ and\ \citenamefont {Guguchia}}]{Das2020}%
  \BibitemOpen
  \bibfield  {author} {\bibinfo {author} {\bibfnamefont {D.}~\bibnamefont
  {Das}}, \bibinfo {author} {\bibfnamefont {K.}~\bibnamefont {Kobayashi}},
  \bibinfo {author} {\bibfnamefont {M.~P.}\ \bibnamefont {Smylie}}, \bibinfo
  {author} {\bibfnamefont {C.}~\bibnamefont {Mielke}}, \bibinfo {author}
  {\bibfnamefont {T.}~\bibnamefont {Takahashi}}, \bibinfo {author}
  {\bibfnamefont {K.}~\bibnamefont {Willa}}, \bibinfo {author} {\bibfnamefont
  {J.-X.}\ \bibnamefont {Yin}}, \bibinfo {author} {\bibfnamefont
  {U.}~\bibnamefont {Welp}}, \bibinfo {author} {\bibfnamefont {M.~Z.}\
  \bibnamefont {Hasan}}, \bibinfo {author} {\bibfnamefont {A.}~\bibnamefont
  {Amato}}, \bibinfo {author} {\bibfnamefont {H.}~\bibnamefont {Luetkens}},\
  and\ \bibinfo {author} {\bibfnamefont {Z.}~\bibnamefont {Guguchia}},\
  }\bibfield  {title} {\bibinfo {title} {Time-reversal invariant and fully
  gapped unconventional superconducting state in the bulk of the topological
  compound ${\mathrm{nb}}_{0.25}{\mathrm{bi}}_{2}{\mathrm{se}}_{3}$},\ }\href
  {https://doi.org/10.1103/PhysRevB.102.134514} {\bibfield  {journal} {\bibinfo
   {journal} {Phys. Rev. B}\ }\textbf {\bibinfo {volume} {102}},\ \bibinfo
  {pages} {134514} (\bibinfo {year} {2020})}\BibitemShut {NoStop}%
\bibitem [{\citenamefont {Fu}\ and\ \citenamefont {Berg}(2010)}]{Fu2010}%
  \BibitemOpen
  \bibfield  {author} {\bibinfo {author} {\bibfnamefont {L.}~\bibnamefont
  {Fu}}\ and\ \bibinfo {author} {\bibfnamefont {E.}~\bibnamefont {Berg}},\
  }\bibfield  {title} {\bibinfo {title} {Odd-parity topological
  superconductors: Theory and application to
  ${\mathrm{cu}}_{x}{\mathrm{bi}}_{2}{\mathrm{se}}_{3}$},\ }\href
  {https://doi.org/10.1103/PhysRevLett.105.097001} {\bibfield  {journal}
  {\bibinfo  {journal} {Phys. Rev. Lett.}\ }\textbf {\bibinfo {volume} {105}},\
  \bibinfo {pages} {097001} (\bibinfo {year} {2010})}\BibitemShut {NoStop}%
\bibitem [{\citenamefont {Fu}(2014)}]{Fu2014}%
  \BibitemOpen
  \bibfield  {author} {\bibinfo {author} {\bibfnamefont {L.}~\bibnamefont
  {Fu}},\ }\bibfield  {title} {\bibinfo {title} {Odd-parity topological
  superconductor with nematic order: Application to
  ${\mathrm{cu}}_{x}{\mathrm{bi}}_{2}{\mathrm{se}}_{3}$},\ }\href
  {https://link.aps.org/doi/10.1103/PhysRevB.90.100509} {\bibfield  {journal}
  {\bibinfo  {journal} {Phys. Rev. B}\ }\textbf {\bibinfo {volume} {90}},\
  \bibinfo {pages} {100509(R)} (\bibinfo {year} {2014})}\BibitemShut {NoStop}%
\bibitem [{\citenamefont {Venderbos}\ \emph
  {et~al.}(2016{\natexlab{a}})\citenamefont {Venderbos}, \citenamefont
  {Kozii},\ and\ \citenamefont {Fu}}]{Venderbos2016}%
  \BibitemOpen
  \bibfield  {author} {\bibinfo {author} {\bibfnamefont {J.~W.~F.}\
  \bibnamefont {Venderbos}}, \bibinfo {author} {\bibfnamefont {V.}~\bibnamefont
  {Kozii}},\ and\ \bibinfo {author} {\bibfnamefont {L.}~\bibnamefont {Fu}},\
  }\bibfield  {title} {\bibinfo {title} {Identification of nematic
  superconductivity from the upper critical field},\ }\href
  {https://doi.org/10.1103/PhysRevB.94.094522} {\bibfield  {journal} {\bibinfo
  {journal} {Phys. Rev. B}\ }\textbf {\bibinfo {volume} {94}},\ \bibinfo
  {pages} {094522} (\bibinfo {year} {2016}{\natexlab{a}})}\BibitemShut
  {NoStop}%
\bibitem [{\citenamefont {Yonezawa}(2018)}]{Yonezawa2018}%
  \BibitemOpen
  \bibfield  {author} {\bibinfo {author} {\bibfnamefont {S.}~\bibnamefont
  {Yonezawa}},\ }\bibfield  {title} {\bibinfo {title} {Nematic
  superconductivity in doped bi2se3 topological superconductors},\ }\href
  {https://doi.org/10.3390/condmat4010002} {\bibfield  {journal} {\bibinfo
  {journal} {Condensed Matter}\ }\textbf {\bibinfo {volume} {4}},\ \bibinfo
  {pages} {2} (\bibinfo {year} {2018})}\BibitemShut {NoStop}%
\bibitem [{\citenamefont {Wu}\ and\ \citenamefont {Martin}(2017)}]{Wu2017}%
  \BibitemOpen
  \bibfield  {author} {\bibinfo {author} {\bibfnamefont {F.}~\bibnamefont
  {Wu}}\ and\ \bibinfo {author} {\bibfnamefont {I.}~\bibnamefont {Martin}},\
  }\bibfield  {title} {\bibinfo {title} {Majorana kramers pair in a nematic
  vortex},\ }\href {https://doi.org/10.1103/PhysRevB.95.224503} {\bibfield
  {journal} {\bibinfo  {journal} {Phys. Rev. B}\ }\textbf {\bibinfo {volume}
  {95}},\ \bibinfo {pages} {224503} (\bibinfo {year} {2017})}\BibitemShut
  {NoStop}%
\bibitem [{\citenamefont {Akzyanov}\ and\ \citenamefont
  {Rakhmanov}(2021)}]{Akzyanov2021a}%
  \BibitemOpen
  \bibfield  {author} {\bibinfo {author} {\bibfnamefont {R.~S.}\ \bibnamefont
  {Akzyanov}}\ and\ \bibinfo {author} {\bibfnamefont {A.~L.}\ \bibnamefont
  {Rakhmanov}},\ }\bibfield  {title} {\bibinfo {title} {Strain-induced spin
  vortex and majorana kramers pairs in doped topological insulators with
  nematic superconductivity},\ }\href
  {https://doi.org/10.1103/PhysRevB.104.094511} {\bibfield  {journal} {\bibinfo
   {journal} {Phys. Rev. B}\ }\textbf {\bibinfo {volume} {104}},\ \bibinfo
  {pages} {094511} (\bibinfo {year} {2021})}\BibitemShut {NoStop}%
\bibitem [{\citenamefont {Hecker}\ and\ \citenamefont
  {Schmalian}(2017)}]{Hecker2018}%
  \BibitemOpen
  \bibfield  {author} {\bibinfo {author} {\bibfnamefont {M.}~\bibnamefont
  {Hecker}}\ and\ \bibinfo {author} {\bibfnamefont {J.}~\bibnamefont
  {Schmalian}},\ }\bibfield  {title} {\bibinfo {title} {Vestigial nematic order
  and superconductivity in the doped topological insulator
  cu$_x$bi$_2$se$_3$},\ }\href
  {https://www.nature.com/articles/s41535-018-0098-z.pdf} {\bibfield  {journal}
  {\bibinfo  {journal} {npj Quantum Mater.}\ }\textbf {\bibinfo {volume} {3}},\
  \bibinfo {pages} {26} (\bibinfo {year} {2017})}\BibitemShut {NoStop}%
\bibitem [{\citenamefont {Hao}\ and\ \citenamefont {Ting}(2017)}]{Hao2017}%
  \BibitemOpen
  \bibfield  {author} {\bibinfo {author} {\bibfnamefont {L.}~\bibnamefont
  {Hao}}\ and\ \bibinfo {author} {\bibfnamefont {C.~S.}\ \bibnamefont {Ting}},\
  }\bibfield  {title} {\bibinfo {title} {Nematic superconductivity in
  ${\mathrm{cu}}_{x}{\mathrm{bi}}_{2}{\mathrm{se}}_{3}$: Surface andreev bound
  states},\ }\href {https://doi.org/10.1103/PhysRevB.96.144512} {\bibfield
  {journal} {\bibinfo  {journal} {Phys. Rev. B}\ }\textbf {\bibinfo {volume}
  {96}},\ \bibinfo {pages} {144512} (\bibinfo {year} {2017})}\BibitemShut
  {NoStop}%
\bibitem [{\citenamefont {Uematsu}\ \emph {et~al.}(2019)\citenamefont
  {Uematsu}, \citenamefont {Mizushima}, \citenamefont {Tsuruta}, \citenamefont
  {Fujimoto},\ and\ \citenamefont {Sauls}}]{Uematsu2019}%
  \BibitemOpen
  \bibfield  {author} {\bibinfo {author} {\bibfnamefont {H.}~\bibnamefont
  {Uematsu}}, \bibinfo {author} {\bibfnamefont {T.}~\bibnamefont {Mizushima}},
  \bibinfo {author} {\bibfnamefont {A.}~\bibnamefont {Tsuruta}}, \bibinfo
  {author} {\bibfnamefont {S.}~\bibnamefont {Fujimoto}},\ and\ \bibinfo
  {author} {\bibfnamefont {J.~A.}\ \bibnamefont {Sauls}},\ }\bibfield  {title}
  {\bibinfo {title} {Chiral higgs mode in nematic superconductors},\ }\href
  {https://doi.org/10.1103/PhysRevLett.123.237001} {\bibfield  {journal}
  {\bibinfo  {journal} {Phys. Rev. Lett.}\ }\textbf {\bibinfo {volume} {123}},\
  \bibinfo {pages} {237001} (\bibinfo {year} {2019})}\BibitemShut {NoStop}%
\bibitem [{\citenamefont {Akzyanov}\ \emph {et~al.}(2020)\citenamefont
  {Akzyanov}, \citenamefont {Kapranov},\ and\ \citenamefont
  {Rakhmanov}}]{Akzyanov2020_2}%
  \BibitemOpen
  \bibfield  {author} {\bibinfo {author} {\bibfnamefont {R.~S.}\ \bibnamefont
  {Akzyanov}}, \bibinfo {author} {\bibfnamefont {A.~V.}\ \bibnamefont
  {Kapranov}},\ and\ \bibinfo {author} {\bibfnamefont {A.~L.}\ \bibnamefont
  {Rakhmanov}},\ }\bibfield  {title} {\bibinfo {title} {Spontaneous strain and
  magnetization in doped topological insulators with nematic and chiral
  superconductivity},\ }\href {https://doi.org/10.1103/PhysRevB.102.100505}
  {\bibfield  {journal} {\bibinfo  {journal} {Phys. Rev. B}\ }\textbf {\bibinfo
  {volume} {102}},\ \bibinfo {pages} {100505(R)} (\bibinfo {year}
  {2020})}\BibitemShut {NoStop}%
\bibitem [{\citenamefont {Khokhlov}\ and\ \citenamefont
  {Akzyanov}(2021{\natexlab{a}})}]{Khokhlov2021a}%
  \BibitemOpen
  \bibfield  {author} {\bibinfo {author} {\bibfnamefont {D.~A.}\ \bibnamefont
  {Khokhlov}}\ and\ \bibinfo {author} {\bibfnamefont {R.~S.}\ \bibnamefont
  {Akzyanov}},\ }\bibfield  {title} {\bibinfo {title} {Pauli paramagnetism of
  triplet cooper pairs in a nematic superconductor},\ }\href
  {https://doi.org/10.1103/PhysRevB.104.214514} {\bibfield  {journal} {\bibinfo
   {journal} {Phys. Rev. B}\ }\textbf {\bibinfo {volume} {104}},\ \bibinfo
  {pages} {214514} (\bibinfo {year} {2021}{\natexlab{a}})}\BibitemShut
  {NoStop}%
\bibitem [{\citenamefont {Chen}\ \emph {et~al.}(2018)\citenamefont {Chen},
  \citenamefont {Chen}, \citenamefont {Yang}, \citenamefont {Du},\ and\
  \citenamefont {Wen}}]{Chen2018}%
  \BibitemOpen
  \bibfield  {author} {\bibinfo {author} {\bibfnamefont {M.}~\bibnamefont
  {Chen}}, \bibinfo {author} {\bibfnamefont {X.}~\bibnamefont {Chen}}, \bibinfo
  {author} {\bibfnamefont {H.}~\bibnamefont {Yang}}, \bibinfo {author}
  {\bibfnamefont {Z.}~\bibnamefont {Du}},\ and\ \bibinfo {author}
  {\bibfnamefont {H.-H.}\ \bibnamefont {Wen}},\ }\bibfield  {title} {\bibinfo
  {title} {Superconductivity with twofold symmetry in
  \text{Bi}$_2$\text{Te}$_3$/\text{FeTe}$_{0.55}$\text{Se}$_{0.45}$
  heterostructures},\ }\href {https://doi.org/10.1126/sciadv.aat1084}
  {\bibfield  {journal} {\bibinfo  {journal} {Science Advances}\ }\textbf
  {\bibinfo {volume} {4}},\ \bibinfo {pages} {eaat1084} (\bibinfo {year}
  {2018})}\BibitemShut {NoStop}%
\bibitem [{\citenamefont {Khokhlov}\ and\ \citenamefont
  {Akzyanov}(2021{\natexlab{b}})}]{Khokhlov2021}%
  \BibitemOpen
  \bibfield  {author} {\bibinfo {author} {\bibfnamefont {D.~A.}\ \bibnamefont
  {Khokhlov}}\ and\ \bibinfo {author} {\bibfnamefont {R.~S.}\ \bibnamefont
  {Akzyanov}},\ }\bibfield  {title} {\bibinfo {title} {Quasiparticle
  interference in doped topological insulators with nematic
  superconductivity},\ }\href {https://doi.org/10.1016/j.physe.2021.114800}
  {\bibfield  {journal} {\bibinfo  {journal} {Physica E: Low-dimensional
  Systems and Nanostructures}\ }\textbf {\bibinfo {volume} {133}},\ \bibinfo
  {pages} {114800} (\bibinfo {year} {2021}{\natexlab{b}})}\BibitemShut
  {NoStop}%
\bibitem [{\citenamefont {Zyuzin}\ \emph {et~al.}(2017)\citenamefont {Zyuzin},
  \citenamefont {Garaud},\ and\ \citenamefont {Babaev}}]{Zyuzin2017}%
  \BibitemOpen
  \bibfield  {author} {\bibinfo {author} {\bibfnamefont {A.~A.}\ \bibnamefont
  {Zyuzin}}, \bibinfo {author} {\bibfnamefont {J.}~\bibnamefont {Garaud}},\
  and\ \bibinfo {author} {\bibfnamefont {E.}~\bibnamefont {Babaev}},\
  }\bibfield  {title} {\bibinfo {title} {Nematic skyrmions in odd-parity
  superconductors},\ }\href {https://doi.org/10.1103/PhysRevLett.119.167001}
  {\bibfield  {journal} {\bibinfo  {journal} {Phys. Rev. Lett.}\ }\textbf
  {\bibinfo {volume} {119}},\ \bibinfo {pages} {167001} (\bibinfo {year}
  {2017})}\BibitemShut {NoStop}%
\bibitem [{\citenamefont {Schmidt}\ \emph {et~al.}(2020)\citenamefont
  {Schmidt}, \citenamefont {Parhizgar},\ and\ \citenamefont
  {Black-Schaffer}}]{Schmidt2020}%
  \BibitemOpen
  \bibfield  {author} {\bibinfo {author} {\bibfnamefont {J.}~\bibnamefont
  {Schmidt}}, \bibinfo {author} {\bibfnamefont {F.}~\bibnamefont {Parhizgar}},\
  and\ \bibinfo {author} {\bibfnamefont {A.~M.}\ \bibnamefont
  {Black-Schaffer}},\ }\bibfield  {title} {\bibinfo {title} {Odd-frequency
  superconductivity and meissner effect in the doped topological insulator
  ${\mathrm{bi}}_{2}{\mathrm{se}}_{3}$},\ }\href
  {https://doi.org/10.1103/PhysRevB.101.180512} {\bibfield  {journal} {\bibinfo
   {journal} {Phys. Rev. B}\ }\textbf {\bibinfo {volume} {101}},\ \bibinfo
  {pages} {180512(R)} (\bibinfo {year} {2020})}\BibitemShut {NoStop}%
\bibitem [{\citenamefont {Akzyanov}(2021)}]{Akzyanov2021b}%
  \BibitemOpen
  \bibfield  {author} {\bibinfo {author} {\bibfnamefont {R.~S.}\ \bibnamefont
  {Akzyanov}},\ }\bibfield  {title} {\bibinfo {title} {Lifshitz transition in
  dirty doped topological insulator with nematic superconductivity},\ }\href
  {https://doi.org/10.1103/PhysRevB.104.224502} {\bibfield  {journal} {\bibinfo
   {journal} {Phys. Rev. B}\ }\textbf {\bibinfo {volume} {104}},\ \bibinfo
  {pages} {224502} (\bibinfo {year} {2021})}\BibitemShut {NoStop}%
\bibitem [{\citenamefont {Buzdin}(2008)}]{Buzdin2008}%
  \BibitemOpen
  \bibfield  {author} {\bibinfo {author} {\bibfnamefont {A.}~\bibnamefont
  {Buzdin}},\ }\bibfield  {title} {\bibinfo {title} {Direct coupling between
  magnetism and superconducting current in the josephson
  ${\ensuremath{\varphi}}_{0}$ junction},\ }\href
  {https://doi.org/10.1103/PhysRevLett.101.107005} {\bibfield  {journal}
  {\bibinfo  {journal} {Phys. Rev. Lett.}\ }\textbf {\bibinfo {volume} {101}},\
  \bibinfo {pages} {107005} (\bibinfo {year} {2008})}\BibitemShut {NoStop}%
\bibitem [{\citenamefont {Sasaki}\ \emph {et~al.}(2020)\citenamefont {Sasaki},
  \citenamefont {Ikegaya}, \citenamefont {Habe}, \citenamefont {Golubov},\ and\
  \citenamefont {Asano}}]{Sasaki2020}%
  \BibitemOpen
  \bibfield  {author} {\bibinfo {author} {\bibfnamefont {A.}~\bibnamefont
  {Sasaki}}, \bibinfo {author} {\bibfnamefont {S.}~\bibnamefont {Ikegaya}},
  \bibinfo {author} {\bibfnamefont {T.}~\bibnamefont {Habe}}, \bibinfo {author}
  {\bibfnamefont {A.~A.}\ \bibnamefont {Golubov}},\ and\ \bibinfo {author}
  {\bibfnamefont {Y.}~\bibnamefont {Asano}},\ }\bibfield  {title} {\bibinfo
  {title} {Josephson effect in two-band superconductors},\ }\href@noop {}
  {\bibfield  {journal} {\bibinfo  {journal} {Physical Review B}\ }\textbf
  {\bibinfo {volume} {101}},\ \bibinfo {pages} {184501} (\bibinfo {year}
  {2020})}\BibitemShut {NoStop}%
\bibitem [{\citenamefont {Kogan}(1981)}]{Kogan1981}%
  \BibitemOpen
  \bibfield  {author} {\bibinfo {author} {\bibfnamefont {V.~G.}\ \bibnamefont
  {Kogan}},\ }\bibfield  {title} {\bibinfo {title} {London approach to
  anisotropic type-ii superconductors},\ }\href
  {https://doi.org/10.1103/PhysRevB.24.1572} {\bibfield  {journal} {\bibinfo
  {journal} {Phys. Rev. B}\ }\textbf {\bibinfo {volume} {24}},\ \bibinfo
  {pages} {1572} (\bibinfo {year} {1981})}\BibitemShut {NoStop}%
\bibitem [{\citenamefont {Ferrell}\ and\ \citenamefont
  {Prange}(1963)}]{Ferrell1963}%
  \BibitemOpen
  \bibfield  {author} {\bibinfo {author} {\bibfnamefont {R.~A.}\ \bibnamefont
  {Ferrell}}\ and\ \bibinfo {author} {\bibfnamefont {R.~E.}\ \bibnamefont
  {Prange}},\ }\bibfield  {title} {\bibinfo {title} {Self-field limiting of
  josephson tunneling of superconducting electron pairs},\ }\href
  {https://doi.org/10.1103/PhysRevLett.10.479} {\bibfield  {journal} {\bibinfo
  {journal} {Phys. Rev. Lett.}\ }\textbf {\bibinfo {volume} {10}},\ \bibinfo
  {pages} {479} (\bibinfo {year} {1963})}\BibitemShut {NoStop}%
\bibitem [{\citenamefont {Kulik}\ and\ \citenamefont
  {Yanson}(1972)}]{Kulik1972}%
  \BibitemOpen
  \bibfield  {author} {\bibinfo {author} {\bibfnamefont {I.~O.}\ \bibnamefont
  {Kulik}}\ and\ \bibinfo {author} {\bibfnamefont {I.~K.}\ \bibnamefont
  {Yanson}},\ }\href@noop {} {\emph {\bibinfo {title} {Josephson Effect in
  Superconducting Tunneling Structures}}}\ (\bibinfo  {publisher} {John Wiley
  and Sons, Inc., New York},\ \bibinfo {year} {1972})\BibitemShut {NoStop}%
\bibitem [{\citenamefont {Kostylev}\ \emph {et~al.}(2020)\citenamefont
  {Kostylev}, \citenamefont {Yonezawa}, \citenamefont {Wang}, \citenamefont
  {Ando},\ and\ \citenamefont {Maeno}}]{Kostylev2020}%
  \BibitemOpen
  \bibfield  {author} {\bibinfo {author} {\bibfnamefont {I.}~\bibnamefont
  {Kostylev}}, \bibinfo {author} {\bibfnamefont {S.}~\bibnamefont {Yonezawa}},
  \bibinfo {author} {\bibfnamefont {Z.}~\bibnamefont {Wang}}, \bibinfo {author}
  {\bibfnamefont {Y.}~\bibnamefont {Ando}},\ and\ \bibinfo {author}
  {\bibfnamefont {Y.}~\bibnamefont {Maeno}},\ }\bibfield  {title} {\bibinfo
  {title} {Uniaxial-strain control of nematic superconductivity in srxbi2se3},\
  }\href {https://doi.org/10.1038/s41467-020-17913-y} {\bibfield  {journal}
  {\bibinfo  {journal} {Nature Communications}\ }\textbf {\bibinfo {volume}
  {11}},\ \bibinfo {pages} {4152} (\bibinfo {year} {2020})}\BibitemShut
  {NoStop}%
\bibitem [{\citenamefont {Venderbos}\ \emph
  {et~al.}(2016{\natexlab{b}})\citenamefont {Venderbos}, \citenamefont
  {Kozii},\ and\ \citenamefont {Fu}}]{Venderbos2016_2}%
  \BibitemOpen
  \bibfield  {author} {\bibinfo {author} {\bibfnamefont {J.~W.~F.}\
  \bibnamefont {Venderbos}}, \bibinfo {author} {\bibfnamefont {V.}~\bibnamefont
  {Kozii}},\ and\ \bibinfo {author} {\bibfnamefont {L.}~\bibnamefont {Fu}},\
  }\bibfield  {title} {\bibinfo {title} {Odd-parity superconductors with
  two-component order parameters: Nematic and chiral, full gap, and majorana
  node},\ }\href {https://doi.org/10.1103/PhysRevB.94.180504} {\bibfield
  {journal} {\bibinfo  {journal} {Phys. Rev. B}\ }\textbf {\bibinfo {volume}
  {94}},\ \bibinfo {pages} {180504(R)} (\bibinfo {year}
  {2016}{\natexlab{b}})}\BibitemShut {NoStop}%
\bibitem [{\citenamefont {Chirolli}(2018)}]{Chirolli2018}%
  \BibitemOpen
  \bibfield  {author} {\bibinfo {author} {\bibfnamefont {L.}~\bibnamefont
  {Chirolli}},\ }\bibfield  {title} {\bibinfo {title} {Chiral superconductivity
  in thin films of doped ${\mathrm{bi}}_{2}{\mathrm{se}}_{3}$},\ }\href
  {https://doi.org/10.1103/PhysRevB.98.014505} {\bibfield  {journal} {\bibinfo
  {journal} {Phys. Rev. B}\ }\textbf {\bibinfo {volume} {98}},\ \bibinfo
  {pages} {014505} (\bibinfo {year} {2018})}\BibitemShut {NoStop}%
\bibitem [{\citenamefont {Chirolli}\ \emph {et~al.}(2017)\citenamefont
  {Chirolli}, \citenamefont {de~Juan},\ and\ \citenamefont
  {Guinea}}]{Chirolli2017}%
  \BibitemOpen
  \bibfield  {author} {\bibinfo {author} {\bibfnamefont {L.}~\bibnamefont
  {Chirolli}}, \bibinfo {author} {\bibfnamefont {F.}~\bibnamefont {de~Juan}},\
  and\ \bibinfo {author} {\bibfnamefont {F.}~\bibnamefont {Guinea}},\
  }\bibfield  {title} {\bibinfo {title} {Time-reversal and rotation symmetry
  breaking superconductivity in dirac materials},\ }\href
  {https://doi.org/10.1103/PhysRevB.95.201110} {\bibfield  {journal} {\bibinfo
  {journal} {Phys. Rev. B}\ }\textbf {\bibinfo {volume} {95}},\ \bibinfo
  {pages} {201110(R)} (\bibinfo {year} {2017})}\BibitemShut {NoStop}%
\bibitem [{\citenamefont {Yuan}\ \emph {et~al.}(2017)\citenamefont {Yuan},
  \citenamefont {He},\ and\ \citenamefont {Law}}]{Yuan2017}%
  \BibitemOpen
  \bibfield  {author} {\bibinfo {author} {\bibfnamefont {N.~F.~Q.}\
  \bibnamefont {Yuan}}, \bibinfo {author} {\bibfnamefont {W.-Y.}\ \bibnamefont
  {He}},\ and\ \bibinfo {author} {\bibfnamefont {K.~T.}\ \bibnamefont {Law}},\
  }\bibfield  {title} {\bibinfo {title} {Superconductivity-induced
  ferromagnetism and weyl superconductivity in nb-doped
  ${\mathbf{bi}}_{2}{\mathbf{se}}_{3}$},\ }\href
  {https://doi.org/10.1103/PhysRevB.95.201109} {\bibfield  {journal} {\bibinfo
  {journal} {Phys. Rev. B}\ }\textbf {\bibinfo {volume} {95}},\ \bibinfo
  {pages} {201109(R)} (\bibinfo {year} {2017})}\BibitemShut {NoStop}%
\bibitem [{\citenamefont {Furusaki}\ \emph {et~al.}(2001)\citenamefont
  {Furusaki}, \citenamefont {Matsumoto},\ and\ \citenamefont
  {Sigrist}}]{Furusaki2001}%
  \BibitemOpen
  \bibfield  {author} {\bibinfo {author} {\bibfnamefont {A.}~\bibnamefont
  {Furusaki}}, \bibinfo {author} {\bibfnamefont {M.}~\bibnamefont
  {Matsumoto}},\ and\ \bibinfo {author} {\bibfnamefont {M.}~\bibnamefont
  {Sigrist}},\ }\bibfield  {title} {\bibinfo {title} {Spontaneous hall effect
  in a chiral p-wave superconductor},\ }\href
  {https://doi.org/10.1103/PhysRevB.64.054514} {\bibfield  {journal} {\bibinfo
  {journal} {Phys. Rev. B}\ }\textbf {\bibinfo {volume} {64}},\ \bibinfo
  {pages} {054514} (\bibinfo {year} {2001})}\BibitemShut {NoStop}%
\bibitem [{\citenamefont {Nambu}(1960)}]{Nambu1960}%
  \BibitemOpen
  \bibfield  {author} {\bibinfo {author} {\bibfnamefont {Y.}~\bibnamefont
  {Nambu}},\ }\bibfield  {title} {\bibinfo {title} {Quasi-particles and gauge
  invariance in the theory of superconductivity},\ }\href
  {https://doi.org/10.1103/PhysRev.117.648} {\bibfield  {journal} {\bibinfo
  {journal} {Phys. Rev.}\ }\textbf {\bibinfo {volume} {117}},\ \bibinfo {pages}
  {648} (\bibinfo {year} {1960})}\BibitemShut {NoStop}%
\bibitem [{\citenamefont {Schrieffer}(1964)}]{schriefferbook}%
  \BibitemOpen
  \bibfield  {author} {\bibinfo {author} {\bibfnamefont {J.~R.}\ \bibnamefont
  {Schrieffer}},\ }\href@noop {} {\emph {\bibinfo {title} {Theory of
  Superconductivity}}}\ (\bibinfo  {publisher} {W.A. Benjamin, San Francisco},\
  \bibinfo {year} {1964})\BibitemShut {NoStop}%
\bibitem [{\citenamefont {Stone}\ and\ \citenamefont {Roy}(2004)}]{Stone2004}%
  \BibitemOpen
  \bibfield  {author} {\bibinfo {author} {\bibfnamefont {M.}~\bibnamefont
  {Stone}}\ and\ \bibinfo {author} {\bibfnamefont {R.}~\bibnamefont {Roy}},\
  }\bibfield  {title} {\bibinfo {title} {Edge modes, edge currents, and gauge
  invariance in ${p}_{x}{+ip}_{y}$ superfluids and superconductors},\ }\href
  {https://doi.org/10.1103/PhysRevB.69.184511} {\bibfield  {journal} {\bibinfo
  {journal} {Phys. Rev. B}\ }\textbf {\bibinfo {volume} {69}},\ \bibinfo
  {pages} {184511} (\bibinfo {year} {2004})}\BibitemShut {NoStop}%
\end{thebibliography}%

\end{document}